\documentclass[10pt]{article}%
\usepackage{amsfonts}
\usepackage{physics}
\usepackage{multirow}
\usepackage{ulem}
\usepackage{bm}
\usepackage{bbm}
\usepackage{subfiles}
\usepackage{amsmath}
\usepackage{graphicx}
\usepackage{hyperref}
\usepackage{verbatim}
\usepackage{color}
\usepackage{tikz}
\usepackage{tikz-3dplot}
\usepackage{pgfplots}
\usepgfplotslibrary{fillbetween}
\usetikzlibrary{patterns}
\usepackage{enumerate}
\usepackage{amsfonts}
\usepackage{amssymb}
\usepackage{authblk}
\usepackage[hang,flushmargin]{footmisc}
\usepackage{lipsum}
\usepackage[font=footnotesize]{caption}
\usepackage{subcaption}
\usepackage[utf8]{} 
\usepackage{empheq}

\csname @addtoreset\endcsname{equation}{section}
\newcommand{\eqnode}[2]{\shortstack{#1\\
		{%
			$\begin{aligned}
				#2
			\end{aligned}$	}}}
\newcommand{\badat}{\begin{alignedat}}
	\newcommand{\eadat}{\end{alignedat}}

\usepackage{csquotes}
\usepackage[backend=biber, hyperref,backref,backrefstyle=none,style=numeric-comp, sorting=none,maxbibnames=5,url=false]{biblatex}

\DefineBibliographyStrings{english}{%
	backrefpage = {cited on page},%
	backrefpages = {cited on pages}%
}
\DeclareFieldFormat{doi}{%
	\hfil\penalty90\hfilneg\space DOI\addcolon\addnbspace
	\ifhyperref
	{\href{https://doi.org/#1}{\nolinkurl{#1}}}
	{\nolinkurl{#1}}}

\DeclareUnicodeCharacter{0301}{*************************************}

\def\be{\begin{eqnarray}}
	\def\ee{\end{eqnarray}}
\def\beann{\begin{eqnarray*}}
	\def\eeann{\end{eqnarray*}}
\def\beq{\begin{equation}}
	\def\eeq{\end{equation}}
\def\ba{\begin{array}}
	\def\ea{\end{array}}
\def\ben{\begin{enumerate}}
	\def\een{\end{enumerate}}
\def\bea{\begin{eqnarray}}
	\def\eea{\end{eqnarray}}

\def\5{\bar }
\def\6{\partial }
\def\7{\hat }
\def\4{\tilde }

\def\cE{\mathcal{E}}

\newcommand{\nspace}{\mkern-18mu}

\newcommand{\st}{S_{\text{st}}}

\usetikzlibrary{arrows, shapes.misc}
\usetikzlibrary{arrows.meta, positioning}
\usetikzlibrary{decorations.markings}
\usetikzlibrary{decorations.pathmorphing}
\usetikzlibrary{backgrounds,automata}
\pgfplotsset{compat=newest}

\tikzset{every picture/.style={line width=0.75pt}} 

\tikzset{cross/.style={cross out, draw=black, minimum size=2*(#1-\pgflinewidth), inner sep=0pt, outer sep=0pt},
	cross/.default={1pt},decoration=
	{markings,
		mark=at position 0.5 with {\arrow{stealth}}
	},
	plain/.style={line width=0.8pt},
	arrow/.style={plain,postaction=decorate}}

\renewcommand{\tilde}{\widetilde}
\renewcommand{\hat}{\widehat}

\begin{document}

\title{\vspace{-70pt} \Large{\sc Reverse stealth construction and its thermodynamic imprints}\vspace{10pt}}
\author[a,b]{\normalsize{Cristi\'an Erices}\footnote{\href{mailto:cristian.erices@ucentral.cl}{cristian.erices@ucentral.cl}}}
\author[c]{\normalsize{Luis Guajardo}\footnote{\href{mailto:luis.guajardo.r@gmail.com}{luis.guajardo.r@gmail.com}}}
\author[d]{\normalsize{Kristiansen Lara}\footnote{\href{mailto:klara@cecs.cl}{klara@cecs.cl}}}
\affil[a]{\footnotesize\textit{Vicerrector\'ia Acad\'emica, Universidad Central de Chile, Toesca $1783$, Santiago $8320000$, Chile}}
\affil[b]{\footnotesize\textit{Departamento de Matemática, Física y Estadística, Universidad Católica del Maule, Av. San Miguel 3605, Talca 3480094, Chile}}
\affil[c]{\footnotesize\textit{Instituto de Investigaci\'on Interdisciplinario, Vicerrector\'ia Acad\'emica, Universidad de Talca, 3465548 Talca, Chile.}}
\affil[d]{\footnotesize\textit{Centro de Estudios Científicos (CECs), Av. Arturo Prat 514, Valdivia, Chile.}}
\date{}
\maketitle
\thispagestyle{empty}

\begin{abstract}
We study a class of solutions within the context of modified gravity theories, characterized by a non-trivial field that does not generate any back-reaction on the metric. These stealth configurations are effectively defined by the stealth conditions, which correspond to a vanishing stress-energy tensor. In this work, we introduce a novel approach to constructing this class of solutions. In contrast to the standard procedure, the starting point requires satisfying the stealth conditions for a given ansatz independently of the gravitational dynamics. This approach simultaneously determines the non-trivial field and the geometries capable of supporting it as a stealth configuration. Consequently, a gravity model can accommodate a stealth field only if its vacuum solution falls within the geometries permissible under stealth conditions. By applying this reverse procedure in the non-minimal $R\phi^2$ coupling, we recover all previously known stealth configurations and present new solutions. Although it seems intuitive to assume that this ``gravitationally undetectable'' scalar field leaves no physical traces, it remarkably reveals thermodynamic imprints, as its presence screens the black hole mass and modifies the entropy according to the first law.

\end{abstract}
\newpage
\begin{small}
{\addtolength{\parskip}{-2pt}
\tableofcontents}
\end{small}
\thispagestyle{empty}
\newpage
\section{Introduction}\label{Section: Introduction}
The no-hair conjecture states that black holes cannot be described by any quantity apart from their mass, electric charge, and angular momentum \cite{wheeler}. This means that in an appropriately chosen reference frame, isolated black holes can be described by the well-known Kerr-Newman solution \cite{kerr,israel,heusler}. According to this conjecture, in the final stage of gravitational collapse, black holes are defined solely by these quantities, which adhere to a Gauss law and are measurable at infinity. This implies that any other information about the matter that falls into a black hole is irretrievably lost. This can be understood from a mathematical perspective; in General Relativity (GR), a set of assumptions renders black hole solutions incompatible with non-trivial matter fields, apart from the electromagnetic one. Specifically, it has been shown that there are no non-trivial regular solutions in GR when minimally coupled scalar fields are considered \cite{janis}. Consequently, under particular conditions that violate these assumptions of the no-hair conjecture, hairy black hole solutions are expected to emerge. One of these assumptions is violated if there is a non-minimal coupling between the scalar field and gravity. This provided one of the earliest examples of a black hole solution with a non-trivial scalar field in an asymptotically flat spacetime more than five decades ago \cite{Bocharova}.

Exploring hairy black holes in scalar-tensor theories has been especially interesting, considering the great interest resurfaced in the community when it was proved that the generalized Galileon theory \cite{Deffayet1,Deffayet2} is equivalent to the Horndeski theory \cite{Kobayashi}. Horndeski theory was proposed fifty years ago as the most general scalar-tensor theory, whose action principle leads to second-order field equations, and consequently, it is free of Ostrogradski instabilities~\cite{Horndeski:1974wa}. Subsequently, the first hairy black hole solution in Horndeski theory~\cite{Rinaldi} was found, and a plethora of hairy black hole solutions were investigated in scalar-tensor theories, motivated by no-hair extensions specially adapted to these theories~\cite{SotiriouBHs,Hui,MaselliBHs}.

However, during the last decade, it was proved that more general scalar-tensor theories can generate higher-order field equations and still avoid Ostrogradski instabilities by adding suitable higher-order modifications. This generalization of Horndeski theory was presented first in Beyond Horndeski and then in Degenerate Higher Order Scalar Tensor (DHOST) theories, providing a rich class of effective field theories~\cite{Crisostomi:2016tcp,Gleyzes:2014dya,Kobayashi:2019hrl,Langlois:2015cwa,Langlois:2017mdk,Langlois:2018dxi}.

Some particularly interesting configurations in modified theories of gravity possess a non-trivial field and do not generate any back-reaction, resulting in a metric that coincides with GR solutions. In scalar-tensor theories, this translates to a vanishing effective stress-energy tensor originating from all the scalar-tensor terms in the action except for the purely metric sector. These solutions, initially introduced in \cite{Ayon-Beato:2004nzi}, are the so-called stealth black holes and represent the simplest example of hairy black holes that are regular at the horizon.

The search for stealth solutions in scalar-tensor theories has a long story. They were introduced almost two decades ago, dubbing this phenomenon a ``Gravitational Cheshire effect" \cite{Ayon-Beato:2005yoq}. Over the years, this became a seminal work in the field, showing that it was possible to dress 
a flat spacetime with a real scalar field that satisfies a nonlinear Klein-Gordon equation without curving spacetime. This configuration is possible due to the non-minimal coupling of the scalar field with the curvature and a self-interacting potential. A footprint of the coupling remains in the stress-energy tensor even when gravity is switched off, and the condition for a vanishing stress-energy tensor fixes the self-interaction potential as a local function of the scalar field depending on two coupling constants. The solutions can describe shock waves and, in the Euclidean continuation, $D$-dimensional instanton configurations.

Stealth black holes with planar topology were obtained in the bi-scalar extension of Horndeski~\cite{Charmousis:2014zaa}. For the shift-symmetric Horndeski theory, it was found that Schwarzschild and a partially self-tuned de-Sitter Schwarzschild black hole are supported by a space and time-dependent scalar field~\cite{Babichev:2013cya}, and a Kerr background can be supported by a static scalar field~\cite{Babichev:2017guv}. Schwarzschild and Kerr backgrounds with non-trivial scalar field were also found in shift-symmetric breaking Horndeski theories~\cite{Motohashi:2018wdq,Minamitsuji:2018vuw}. Stealth scalar fields were described in shift-symmetric scalar-tensor theories of gravity in the family of Degenerate Higher Order Scalar-Tensor (DHOST) theories for Schwarzschild-(A)dS black holes~\cite{BenAchour:2018dap}, rotating backgrounds such as Kerr black holes~\cite{Charmousis:2019vnf} and in cosmological setups~\cite{Ayon-Beato:2015mxf,Ayon-Beato:2013bsa,Maeda:2012tu}. The existence and geometric properties of stealth solutions have been explored in different scenarios other than scalar-tensor theories, such as 3D New Massive Gravity~\cite{Hassaine:2013cma}, hypergravity~\cite{Henneaux:2015tar}, and interacting gauge fields~\cite{Cisterna:2016nwq,Smolic:2017bic,Quinzacara:2019pes}.

To find stealth black holes in those scenarios, one must circumvent the underlying no-hair theorems and additionally ensure that the stress-energy tensor vanishes. In general terms, a scalar-tensor action can be decomposed as
\begin{align}
S[\phi,g]=S_g[g]+\st[\phi,g]\ ,
\end{align}
where $S_g$ is a purely geometric part and consequently depends only on the metric tensor $g_{\mu\nu}$. The last term $\st$ denotes the scalar-tensor part and is the only one that contains the scalar field $\phi$.

A nontrivial\footnote{By nontrivial, we refer to a non-constant scalar field $\phi$.} scalar field $\phi$ defines a stealth solution if the following conditions are satisfied
\begin{subequations}
\label{eq: Stealth condition global}
\begin{align}
\label{eq: metriceq}
\mathcal{E}_{\mu\nu}&=0\ ,\\
\label{eq: Stealth condition}
T_{\mu\nu}&=0\,.
\end{align}
\end{subequations}
Here, we denote by
\begin{align}
\mathcal{E}_{\mu\nu}=\displaystyle\frac{\delta S_g}{\delta g^{\mu\nu}}\ ,
\end{align}
to the Euler-Lagrange equations generated by the purely metric sector $S_g$. It is worth emphasizing that $\cE_{\mu\nu}$ is not necessarily the Einstein tensor $G_{\mu\nu}$, as $S_g$ can contain purely geometric terms other than the Einstein-Hilbert one. On the other hand, the stress-energy tensor associated with the scalar-tensor part is
\begin{align}
T_{\mu\nu}=\displaystyle-\frac{2}{\sqrt{-g}} \frac{\delta \st}{\delta g^{\mu\nu}}\ .
\end{align}
Equations \eqref{eq: Stealth condition global}
 defy Wheeler's paradigm, which states that ``spacetime tells matter how to move; matter tells spacetime how to curve''~\cite{Wheeler:1998vs}. In a broader sense that fits with scalar-tensor theories, the scalar field $\phi$ is generating no backreaction in the spacetime, yielding an effective equation $\mathcal{E}_{\mu\nu}=0$, forgoing the sourcing of the stress-energy tensor $T_{\mu\nu}$.

The usual procedure for the construction of stealth solutions is the following: \textit{a)} For a given spacetime ansatz, one solves the set of field equations \eqref{eq: metriceq} generated from a specific action $S_g$, obtaining the corresponding solution $g_{\mu\nu}$. Then, step \textit{b)} requires solving the stealth conditions \eqref{eq: Stealth condition}, with $g_{\mu\nu}$ as a background, to obtain a nontrivial scalar field. This last step, of course, depends on $\st$ and how the assumptions of the corresponding no-hair conjecture are circumvented. We dub this procedure as the ``$\textit{a)}\rightarrow \textit{b)}$'' method or standard procedure.

This paradigm has been applied extensively, where the standard non-minimal coupling of the form $R \phi^2$ has proven to be highly successful in constructing stealth solutions. It was useful not only for dressing Minkowski spacetime with stealth solutions but also for constructing a stealth scalar field compatible with the non-rotating Bañados-Teitelboim-Zanelli (BTZ) black hole in 2+1 dimensions~\cite{Ayon-Beato:2004nzi}. The black hole was supported by a time-dependent scalar field and a self-interacting potential. These works represent one of the simplest ways to obtain stealth solutions. For this reason, it was exploited systematically in other theories that include this term, finding stealth scalar fields in the particular case with conformal coupling~\cite{Anabalon:2009qt,Cardenas:2014kaa,Aviles:2018vnf}, with the generic non-minimal coupling in Anti-de Sitter (AdS) spacetimes~\cite{Hassaine:2006gz,Caldarelli:2013gqa,Flores-Alfonso:2023fkd}, and even in Lifshitz spacetimes~\cite{Bravo-Gaete:2021kgt}. In higher dimensional scalar-tensor theories, stealth solutions were also found in Einstein-Gauss-Bonnet gravity non-minimally coupled to a scalar field~\cite{BravoGaete:2013acu}, its Lovelock extension~\cite{BravoGaete:2013djh}, and recently in more general contexts~\cite{Babichev:2023rhn,Ayon-Beato:2024xgp}. Besides, stealth configurations over the Riegert black hole were found for conformal gravity with a conformally coupled scalar field in~\cite{Brihaye2009,Anastasiou:2022wjq}. 


Since the standard non-minimal coupling admits different specific geometries, it seems natural to ask whether those spacetimes fit into a more general family of solutions. We have found that, to answer this question, it is necessary to reverse the aforementioned ``$\textit{a)}\rightarrow \textit{b)}$'' method. In this way, our starting point is \textit{b)}, which means to satisfy the stealth conditions \eqref{eq: Stealth condition}, without fixing the background $g_{\mu\nu}$ to any particular solution, or equivalently, without specifying any purely geometric sector $S_g$. Consequently, only the stealth conditions \eqref{eq: Stealth condition} on the scalar field constrain the spacetime geometry. The final step is to find purely gravitational theories compatible with the geometry induced by the stealth conditions. We dub this procedure as the ``$\textit{b)}\rightarrow \textit{a)}$'' method, or reverse procedure.

The reverse procedure in constructing stealth solutions brings new implications that are easy to overlook when the standard procedure is applied. This is because different purely geometric actions can admit the same particular solution for their field equations \eqref{eq: metriceq} and are compatible with the same non-trivial scalar field obtained in \eqref{eq: Stealth condition}. In this sense, the mapping that associates elements from the set of solutions obtained in step \textit{a)} to the elements of the set of solutions obtained in step \textit{b)} is not bijective. This means that solely the stealth conditions determine the family of spacetimes compatible with a non-minimally coupled scalar field. As a consequence, any purely geometric action whose metric solution fits in this family of metrics can be endowed with a stealth scalar field, broadening the landscape of theories that generate non-minimally coupled stealth solutions. Thus, it is possible not only to recover and extend existing solutions in the literature but also to find new solutions by easily identifying the theories that admit stealth configurations.

The generality of the solution obtained in this work puts us in a position to explore its physical consequences. While it is intuitive to think that no physical imprints are left by ``gravitationally undetectable'' objects, such as stealth configurations, unexpectedly, we found that the mass of stealth black holes is shifted compared to its trivial counterpart. The latter phenomenon can be seen explicitly in a concrete example by selecting, as the purely geometric theory, the five-dimensional Einstein-Gauss-Bonnet gravity.

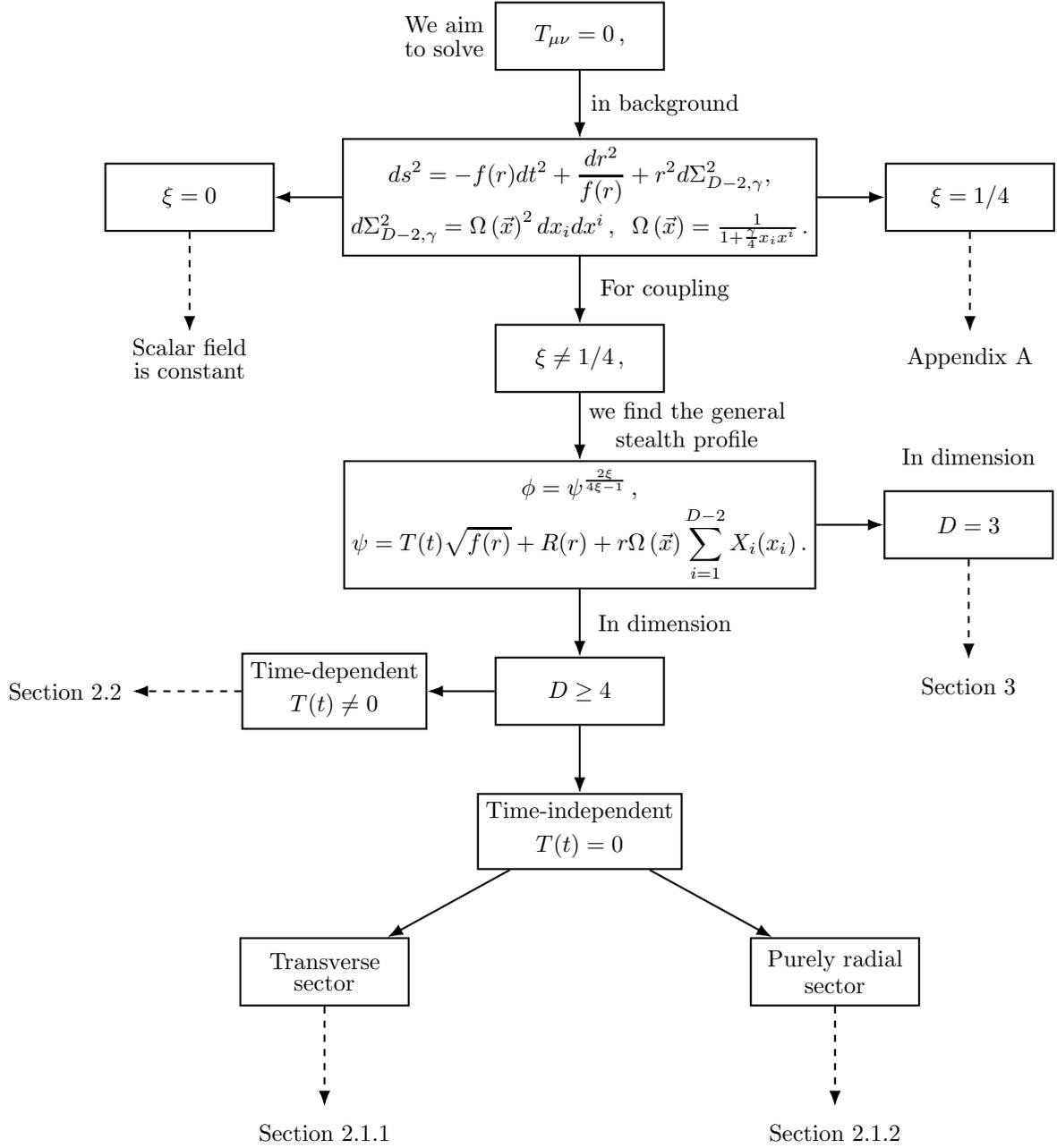
\begin{figure}[h!]
	\centering
 \begin{tikzpicture}[
			every node/.style={rectangle, draw, minimum width=2.5cm, minimum height=1cm},
			every path/.style={-Latex, thick}
			]
			
			\node (Tmunu) at (0,0) {$T_{\mu\nu}=0\,,$};
			\node (metric) [below=of Tmunu] {\shortstack{$\displaystyle ds^2=-f(r)dt^2+\frac{dr^2}{f(r)}+r^2d\Sigma^2_{D-2,\gamma}$,\\
			$d\Sigma^2_{D-2,\gamma}=\Omega\left(\vec{x}\right)^2dx_idx^i\,,\;\; \Omega\left(\vec{x}\right)=\frac{1}{1+\frac{\gamma}{4}x_ix^i}\,.$}};
			\node (xi=0) [left=of metric] {$\xi=0$};
			\node (xi=1/4) [right=of metric] {$\xi=1/4$};

			
			\node (xineq1/4) [below=of metric] {$\xi\neq 1/4\,,$};
			\node (angular solution) [below=of xineq1/4] {\shortstack{$\phi=\psi^{\frac{2\xi}{4\xi-1}}\,,$\\
			$\displaystyle\psi=T(t)\sqrt{f(r)}+R(r)+r\Omega\left(\vec{x}\right)\sum_{i=1}^{D-2}X_i(x_i)\,.$}};
			
			\node (Deq3) [right=of angular solution] {$D=3$};

			\node (Dgeq4) [below=of angular solution] {$D\geq 4$};
			\node (Time-independent Dgeq4) [below=of Dgeq4] {\eqnode{Time-independent}{T(t)=0}};
			\node (Time-dependent Dgeq4) [left=of Dgeq4] {\eqnode{Time-dependent}{T(t)\neq 0}};
			
			\node (Transverse sector) [below left=of Time-independent Dgeq4] {\shortstack{Transverse\\sector}};
			\node (Purely radial sector) [below right=of Time-independent Dgeq4] {\shortstack{Purely radial\\sector}};

			\draw (Tmunu) node[left,draw=none,xshift=-0.75cm] {\shortstack{We aim \\
				to solve}} -- (metric) node[midway,right,draw=none] {\shortstack{in background}};
			\draw (metric) -- (xi=0);
			\draw (metric) -- (xi=1/4);
			\draw (metric) -- (xineq1/4) node [midway,right,draw=none] {For coupling};
			\draw (xineq1/4) -- (angular solution) node[midway, right,draw=none] {\shortstack{we find the general\\stealth profile}};
			\draw (angular solution)-- (Dgeq4) node [above right,draw=none,yshift=0.5cm] {\shortstack{In dimension}};
			\draw (angular solution) -- (Deq3) node [above,draw=none,yshift=0.5cm] {\shortstack{In dimension}};
			\draw (Dgeq4) -- (Time-independent Dgeq4); 
			\draw (Dgeq4)-- (Time-dependent Dgeq4);
			\draw[dashed] (Deq3) -- ++(0,-2) node[draw=none,yshift=-0.4cm] {\shortstack{Section \ref{Section: Three-dimensional case}}};
			\draw (Time-independent Dgeq4)--(Transverse sector);
			\draw (Time-independent Dgeq4)--(Purely radial sector);
			
			\draw[dashed] (xi=0) -- ++(0,-2) node[draw=none,yshift=-0.4cm] {\shortstack{Scalar field\\ is constant}};
			
			\draw[dashed] (xi=1/4) -- ++(0,-2) node[draw=none,yshift=-0.4cm] {\shortstack{Appendix \ref{Appendix: Section: Case xi=1/4}}};
			
			
			
			\draw[dashed] (Time-dependent Dgeq4) -- ++(-3,0) node[draw=none,xshift=-1cm] {\shortstack{Section \ref{Subsection: Time-dependent configuration Dgeq4}}};
			
			
			

			\draw[dashed] (Transverse sector) -- ++(0,-2) node[draw=none,yshift=-0.4cm] {Section \ref{Subsubsection: Transverse sector}};
			
			\draw[dashed] (Purely radial sector) -- ++(0,-2) node[draw=none,yshift=-0.4cm] {Section \ref{Subsubsection: Purely radial sector}};

		\end{tikzpicture}
	\captionof{figure}{Guide route}
    \label{fig: Scheme}
\end{figure}

This article is organized as follows. To address point \textit{b)}, in Section \ref{Section: The reverse stealth construction}, we present the reverse approach in $D\geq 4$ dimensions and construct the stealth solution. We analyze its different branches and discuss their implications. Given its particular nature, Section \ref{Section: Three-dimensional case} is devoted to the three-dimensional case. A scheme of the reverse procedure can be seen in Fig.~\ref{fig: Scheme}. Then, we develop step \textit{a)} in Section \ref{Section: Applying the reverse stealth construction: Concrete examples}. We apply the reverse procedure to different gravitational theories, as seen in Table \ref{tab: Applications}, recovering and generalizing known results from the literature. We also report novel stealth solutions. In Section \ref{Section: Thermodynamic imprints of stealth black holes}, the Euclidean approach is applied to study thermodynamic aspects of stealth black holes in the five-dimensional Einstein-Gauss-Bonnet theory, demonstrating that the stealth scalar field shifts the black hole mass. In Section \ref{Section: Summary and outlook}, we summarize, discuss, and explore potential avenues for further research.

\newpage


\newpage


\section{Constructing stealth configurations: The reverse procedure}\label{Section: The reverse stealth construction}
The reverse approach proposes a generic procedure to construct stealth solutions. However, we must ground this approach in a specific scalar-tensor theory. Accordingly, hereafter, we will focus on the non-minimally coupled scalar field in $D$ dimensions,
\begin{align}
\label{eq: Nonminimal coupling}
\st[\phi,g]=-\int d^Dx\sqrt{-g}\left(\frac{1}{2}\nabla_\mu\phi\nabla^\mu\phi+\frac{1}{2}\xi R\phi^2+U\left(\phi\right)\right)\,,
\end{align}
where $\xi\neq 0$ is the coupling constant and $U(\phi)$ is a self-interacting potential. Inspired by works~\cite{Ayon-Beato:2004nzi,Ayon-Beato:2005yoq,Hassaine:2006gz,Hassaine:2013cma,BravoGaete:2013acu,BravoGaete:2013djh,Babichev:2013cya}, we will only consider static and homogeneous spacetimes,
\begin{align}
\label{eq:metric_ansatz}
\dd s^2=-f(r)\dd t^2+\frac{\dd r^2}{f(r)}+r^2\dd\Sigma^2_{D-2,\gamma}\,,
\end{align}
where $\dd\Sigma_{D-2,\gamma}^2$ denotes the line element of a Euclidean, codimension-two base manifold, locally isometric to the sphere $\mathbb{S}^{D-2}$, flat space $\mathbb{R}^{D-2}$, or to the hyperbolic manifold $\mathbb{H}^{D-2}$, with normalized constant curvature $\gamma=\{+1,0,-1\}$, respectively. Namely, it can be written as,
\begin{align}
\dd\Sigma_{D-2,\gamma}^2 =\Omega\left(\vec{x}\right)^2 \dd x_i \dd x^i\,,\quad \Omega\left(\vec{x}\right)=\frac{1}{1+\frac{\gamma}{4}x_i x^i}\,,
\end{align}
where $x^i$ are the Euclidean coordinates of the base manifold with $i=\{1,\dots,D-2\}$. Any contraction between upper and lower Latin indices is understood to be performed using the Kronecker delta $\delta^{ij}$. If both indexes are at the same level, they are not contracted but fixed.

Here, we construct stealth solutions in $D\geq 4$ dimensions without restricting, in principle, the coordinate dependence of the scalar field. In this context, the Euler-Lagrange equations allow us to define the following stress-energy tensor\footnote{The scalar field equation is satisfied by virtue of the Bianchi identity.}
	\begin{align}
		\label{eq: Tmunu}
		T_{\mu \nu}&=\nabla_{\mu}\phi\nabla_{\nu}\phi - g_{\mu\nu}\left(\frac{1}{2}\nabla_{\sigma}\phi\nabla^{\sigma}\phi +U(\phi) \right) + { \xi\left(g_{\mu\nu}\Box -\nabla_{\mu} \nabla_{\nu}+G_{\mu\nu} \right)\phi^2}\,.
	\end{align} 	
	We aim to solve $T_{\mu\nu}=0$ in the ansatz \eqref{eq:metric_ansatz}. The potential $U\left(\phi\right)$ required to sustain the stealth solution is sensitive to the curvature $\gamma$, so it is kept undetermined at the moment. 
	
	We begin analyzing the off-diagonal components of the stress-energy tensor. For fixed, but distinct indices $\mu$ and $\nu$, it reads
	\begin{align}
		\label{eq: Tij}
		T_{\mu}^{\nu} &= \nabla_{\mu}\phi\nabla^{\nu}\phi - \xi \nabla_{\mu} \nabla^{\nu}\phi^2\,,\quad \mu\neq \nu.
	\end{align}
	The above can be rewritten in appealing form, according to the following change of variables
	\begin{align}
		\label{eq: Phi}
		\phi=\psi^{\frac{2\xi}{4\xi-1}}\,,
	\end{align}
	with $\xi \neq 1/4$, obtaining
	\begin{align}
		\label{eq:T_offdiag}
		T_{\mu}^\nu &= -\dfrac{4\xi^2}{(4\xi-1)}\psi^{\frac{1}{4\xi-1}}\nabla_{\mu}\nabla^\nu \psi = 0\,,\quad \mu\neq \nu\,.
	\end{align}	
	Regarding case $\xi=1/4$, it is treated in Appendix \ref{Appendix: Section: Case xi=1/4}. Therefore, the off-diagonal equations \eqref{eq:T_offdiag} are akin to the sum-separability condition. Because of this, we can obtain a simple generic profile for the scalar field. In terms of the auxiliary function $\psi$, it reads
	\begin{align}
		\label{eq: Scalar_profile}
			\psi(t,r,x_1,\dots,x_{D-2})&=T(t)\sqrt{f(r)}+R(r)+r\Omega\left(\vec{x}\right)\sum_{i=1}^{D-2}X_i(x_i)\,,
	\end{align}
	where $T(t)$, $R(r)$ and $X_i\left(x_i\right)$ are functions that depends only on $t$, $r$ and the base manifold coordinates $x_i$, respectively.
	
Using this expression, the next step is to determine the form of the unknown functions $T(t)$, $R(r)$, and $X_i(x_i)$. Taking the difference $T_{i}^{i} - T_{j}^{j}=0$ for fixed but distinct spatial coordinates $i,j=\{1,\dots\,, D-2\}$ (the components of the difference shall not be confused with the trace), we arrive to
	\begin{align}
		T^{i}_{i}-T^{j}_{j}\propto\dv[2]{X_i}{x_i}-\dv[2]{X_j}{x_j}=0\,,\quad i,j=\{1,\dots,D-2\}\,.
	\end{align}
	The solution is given by
	\begin{align}
		\label{eq: Xi}
		X_i(x_i)=a x_i^2+b_ix_i+c_i\,,\quad i=\{1,2,\dots,D-2\}\,,
	\end{align}
	where $a$, $b_i$ and $c_i$ are integration constants. Thus far, we have $T^{1}_{1}=T^{2}_{2}=\dots=T^{{D-2}}_{{D-2}}$. 
	
	The difference $T^t_t-T^r_r$ yields
	\begin{align}
		\begin{split}
			\label{eq: Ttt-Trr}
			T^t_t-T^r_r&=-\dfrac{\psi^{\frac{1}{4\xi-1}}\xi^2}{\sqrt{f}(4\xi-1)}\left( -4f^{3/2}R'' - 4\ddot{T} - 2Tff''+ Tf^{\prime^2} \right)= 0 \,.
		\end{split}
	\end{align}
	In the above expression, primes and dots denote radial and time derivatives, respectively. 
	
It is necessary to distinguish in \eqref{eq: Ttt-Trr} that the function $T(t)$ can be constant, generating an additional branch of solutions. These can be seen explicitly by differentiating the expression inside the brackets with respect to the time and the radius, leading to the compatibility condition
	\begin{align}
		\label{eq: Compatibility condition}
		2ff'''\dot{T}=0\,.
	\end{align}
The latter admits two nontrivial branches with a time (in)dependent scalar field. After removing redundancies, we get,
\begin{align}\label{eq: Branches}
		R(r)= R_1 r + R_0\,,\quad \begin{cases}T(t)=0&\rightarrow\text{Time-independent scalar field}\\T(t)\neq\text{cst}&\rightarrow\text{Time-dependent scalar field}\end{cases}
\end{align}	
At this stage of the procedure, the stress-energy tensor is diagonal, with $T^{t}_{t} = T^{r}_{r}$, and $T^{1}_{1} = T^{2}_{2} = \dots = T^{D-2}_{{D-2}}$. The next subsection \ref{Subsection: Time-independent configuration Dgeq4} is devoted to developing the time-independent branch of solutions. We must mention that the time-dependent branch in Subsection \ref{Subsection: Time-dependent configuration Dgeq4} has some overlap with an unpublished work\footnote{Ayón-Beato, Martínez, Troncoso and Zanelli, private communication.}.

\subsection{Time-independent configuration}\label{Subsection: Time-independent configuration Dgeq4}

The construction of time-independent stealth solutions will require distinguishing between those with a scalar field depending on the spatial coordinates and those with purely radial dependence. To see this, we focus our attention on the difference $T^t_t-T^j_j$ (do not confuse with the trace), which can be schematically written in the form
\begin{align}
    \label{eq: Ttt-Tii}
	T^t_t-T^{j}_{j}&=\dfrac{\Omega\left(\vec{x}\right)}{8r^2\sqrt{f}(4\xi-1)\psi^{\frac{8\xi-1}{4\xi-1}}}\left( \Xi_2 x_ix^i + \Xi_1 b_ix^i + \Xi_0 \right) = 0\,,
\end{align}
where $\Xi_{2,1,0}$ are differential equations detailed in Eq.~\eqref{ap: eq: Xis} of Appendix \ref{Appendix: Explicit form}. At this point, we see that the time-independent branch provides two sub-branches of solutions:
\begin{align}
\label{eq: subbranches}
b_i &\neq 0\,, & \phi &= \phi(r,\vec{x})\,, & \rightarrow & \quad \text{Spatial coordinates-dependent scalar field}, \\
b_i &= 0\,, & \phi &= \phi(r)\,, & \rightarrow & \quad \text{Radial-dependent scalar field}.
\end{align}
It is worth emphasizing that since the reverse approach does not impose any prior restrictions on the coordinate dependence of the scalar field, the spatial coordinates-dependent scalar field has been overlooked in the literature. This aspect gives rise to novel classes of stealth solutions.

\subsubsection{Spatial-coordinates dependent scalar field}\label{Subsubsection: Transverse sector}
The condition $b_i\neq 0$ leads to solve the set of differential equations $\Xi_{0}=0$, $\Xi_{1}=0$ and $\Xi_{2}=0$. We start with $\Xi_1=0$, which is a differential equation for the metric function $f(r)$, whose solution is
\begin{align}
	\label{eq: Metric function angular}
	f(r)=Ar^2 +Br^{\frac{(D-3) - 4(D-2)\xi}{4\xi-1}} +\gamma\,,
\end{align}
where $A$ and $B$ are integration constants. Replacing in $\Xi_0=0$ and $\Xi_2=0$, the following conditions are obtained,
\begin{align}
B R_0&=0\,,\label{condsB}\\
\gamma\left(2R_1 + \sum_{i=1}^{D-2}c_i \right) + 4a &= 0\,,\label{condsa}
\end{align}
where $a$ and $c_i$ are the integration constants of \eqref{eq: Xi}. Note that equation \eqref{condsB} prevents having non-vanishing constants $B$ and $R_0$ simultaneously. On the other hand, equation \eqref{condsa} allows us to write $a$ in terms of $\gamma$, $R_1$, and $c_i$. Then, the auxiliary scalar field $\psi$ obtained in \eqref{eq: Scalar_profile} lead to the following profile for the scalar field,
\begin{align}\label{scalar_rx}
	\phi(r,\vec{x})&=\left\{R_0+r \left[(2\Omega\left(\vec{x}\right) - 1)\left(R_1 + \sum_{i=1}^{D-2}c_i \right) + \Omega\left(\vec{x}\right) b_ix^i \right]\right\}^\frac{2\xi}{4\xi-1}\,.
\end{align}
At this stage, all the diagonal components of the stress-energy tensor are equal. Finally, the stealth conditions $T_{\mu\nu}=0$ are satisfied by the following self-interacting potential,
\begin{align}
\label{eq: potencial_general_structure}
U(\phi) = \lambda_1 \phi^2 + \lambda_2 \phi^{-\frac{2\xi-1}{\xi}} + \lambda_3\phi^{\frac{1}{2\xi}}\,,
\end{align}
with 
\begin{subequations}
\label{eq: Potencial_general_lambdas}
\begin{align}
\label{eq: lambda1}
&\lambda_1 = \dfrac{\xi AD(D-1)(\xi - \xi_D)(\xi-\xi_{D+1})}{2\left(\xi -1/4\right)^2}\,, \\
\label{eq: lambda2}
&\lambda_2 = \dfrac{\xi^2}{8(\xi - 1/4)^2}\left[ AR_0^2 + \gamma\left(R_1 + \displaystyle \sum_{i=1}^{D-2}c_i\right)^2+\sum_{i=1}^{D-2}b_i^2 \right]\,,\\
\label{eq: lambda3}
 &\lambda_3 = - \dfrac{A R_0 \xi^2 (D-1)(\xi - \xi_D)}{\left(\xi - 1/4\right)^2}\,.
\end{align}
\end{subequations}
In general, since the supporting potential determines the theory given in the action principle, its constants must be considered as parameters of the theory. Therefore, they are fixed without variation, implying a relation between the variations of the integration constants. In particular, we see that \eqref{eq: Potencial_general_lambdas} implies $\delta\lambda_1=\delta\lambda_2=\delta\lambda_3=0$. This feature is relevant for the thermodynamic aspects discussed in Section \ref{Section: Thermodynamic imprints of stealth black holes}. As it is expected, the potential in \eqref{eq: potencial_general_structure}, restore conformal invariance when $\xi = \xi_D = \dfrac{D-2}{4(D-1)}$.

So far, a spatial-coordinate dependent scalar field dictates a family of admissible geometries given by the metric function \eqref{eq: Metric function angular}. The additional restriction $BR_0=0$ \eqref{condsB} preserves this structure if $R_0=0$ and $B\neq0$. Otherwise, if $R_0\neq0$ and $B=0$, the geometry is isometric to (A)dS$_D$, meaning that any purely gravitational theory whose background solution is maximally symmetric admits a non-minimally coupled stealth scalar field.

\subsubsection{Radial-dependent scalar field}\label{Subsubsection: Purely radial sector}

In contrast to the previous subsection, the condition $b_i=0$ implies solving the set of differential equations given by $\Xi_{0}=0$ and $\Xi_{2}=0$. A suitable combination of such equations gives,
\begin{align}
\label{eq: constraint a y c_i}
	\gamma\sum_{i=1}^{D-2}c_i-4a=0\,.
\end{align}
This implies, after absorbing redundant constants, a radial-dependent scalar field, whose form is given by
\begin{align}\label{radial_scalar}
    \phi(r) = (R_1r +R_0)^{\frac{2\xi}{4\xi-1}}\,.
\end{align}
The aforementioned set of differential equations is solved by lengthy expressions for the metric function that depends on Gauss' hypergeometric and Gamma functions. However, familiar solutions are found in the conformal case where $\xi = \xi_D = \frac{D-2}{4(D-1)}$. Namely \footnote{Conformal couplings of different dimensions generate logarithmic contributions in the metric function. We opt not to elaborate further on those cases.},
\begin{align}
\label{eq: f conformal}
f(r) = Ar^2 + 2\gamma \dfrac{R_1}{R_0}r + \gamma + \dfrac{B(R_1r+R_0)^{D-1}}{r^{D-3}}\,,\quad R_0\neq 0\,,
\end{align}
from where we find the conformally-invariant potential
\begin{align}
    \label{eq: U conformal}
    U(\phi) = \lambda\phi^{\frac{2D}{D-2}}\,,\quad\text{with}\quad \lambda= AR_0^2-\gamma R_1^2.
\end{align}
In the conformal case with $R_0=0$, the scalar field gets rigid, i.e., its integration constant $R_1$ is fixed without variation as the potential requires it to be a parameter of the theory. Nevertheless, the stress-energy tensor vanishes for a geometry ruled by $f(r)=Ar^2+Br - 2\gamma$, and a conformally invariant potential.


With the reverse approach, we can confirm that in dimension $D\geq 4$, the time-independent stealths compatible with General Relativity, with or without cosmological constant, are the stealth solutions in Minkowski~\cite{Ayon-Beato:2005yoq} and (A)dS$_D$ spacetimes\footnotemark[1]. Therefore, in this set-up, finding black hole geometries for a given dress necessarily requires going beyond General Relativity (see Section \ref{Section: Applying the reverse stealth construction: Concrete examples} for examples).

\subsection{Time-dependent configuration}\label{Subsection: Time-dependent configuration Dgeq4}

Now, we will turn to some comments about the time-dependent branch. From the compatibility condition \eqref{eq: Compatibility condition}, one can deduce that the metric function is at most a quadratic polynomial in $r$,
\begin{align}
    \label{eq: Metric function time-dependent branch}
    f(r)=Ar^2+Br+C\,.
\end{align}
Plugging the metric function in equations $T^t_t-T^r_r=0$ and $T_{t}^{t} - T_{i}^{i}=0$, determines $C=\gamma$, and the following scalar profile,
\begin{align}
\phi(t,r, \vec{x})=\left\{R_0+T(t) \sqrt{f(r)}+R_1 r+r \Omega(\vec{x}) \sum_{i=1}^{D-2} X_i\left(x_i\right)\right\}^{\frac{2 \xi}{4 \xi-1}}\ ,
\end{align}
where the nature of this time-dependent configuration is provided by $T(t)$ which is given by,
\begin{align}
T(t)=T_1 \cos \left(\frac{\sqrt{-\Delta}}{2} t\right)+\frac{T_0}{\sqrt{-\Delta}} \sin \left(\frac{\sqrt{-\Delta}}{2} t\right)\,,\quad \Delta=B^2-4 A \gamma\,.
\end{align}
A real scalar field requires a positive leading term in the asymptotic behavior of the metric function $f(r)$. This implies, in particular, that asymptotically AdS metrics can be supported by real stealth configurations, unlike their dS counterparts. The solution for Minkowski stealths from Ref.~\cite{Ayon-Beato:2005yoq} is recovered when $f(r)$ is a constant ($A=B=0$ and $\gamma=1$).

The functional form of the self-interacting potential is the same as in the time-dependent configuration \eqref{eq: potencial_general_structure},
\begin{equation}
U(\phi)=\lambda_1 \phi^2+\lambda_2 \phi^{-\frac{2 \xi-1}{\xi}}+\lambda_3 \phi^{\frac{1}{2 \xi}}\,,
\end{equation}
except that the parameters $\lambda_k=\lambda_k\left(A, R_0, R_1, b_j, c_j, T_0, T_1\right)$ with $j=\{1,\dots,D-2\}$, $k=\{1,2,3\}$ obeys new relations with the integration constants. However, these new relations do not give new insights about the solution. It is worth noticing that the case $B=0$ forces $R_1=0$. Otherwise, $\xi$ must be fixed to the conformal value $\xi_D$, and $R_1 = BR_0/2$.

\section{Three-dimensional case} \label{Section: Three-dimensional case}

The reverse approach used in the previous section is slightly different in $D=3$, with $\gamma=0$, mainly because we lack angular components to perform non-trivial subtractions in the spatial sector of $T^\mu_\nu$. This situation will be reflected in the metric function, whose additive constant might differ from the spatial curvature $\gamma$. Apart from this fact, three cases arise depending on the coordinate dependencies for the scalar field, as in the previous section. Again, we focus on homogeneous configurations,
\begin{align*}
\dd s^2=-f(r)\dd t^2+\frac{\dd r^2}{f(r)}+r^2\dd\theta^2\,.
\end{align*}

\paragraph{Radial-dependent scalar field:} In the purely-radial case, we readily check that
\begin{align}
    \label{eq:phi-purely-radial-3d}
    \phi(r)=(R_1r+R_0)^{\frac{2\xi}{4\xi-1}}\,,
\end{align}
while the metric function $f(r)$ is generically written in terms of a hypergeometric function. However, as in the $D\geq 4$ case, some particular choices relax the latter behavior. For example, choosing $R_0=0$ is consistent with taking $D=3$ and $\gamma=0$ in Eq.\eqref{eq: Metric function angular}. The interesting choice is the conformal case with $\xi = 1/8$, where the metric function is a quadratic polynomial,
\begin{align}
    \label{eq:f-purely-radial-xi-conf}
    f(r)=Ar^2 + Br + C\,,
\end{align}
and the constants $B$ and $C$ are tied with the constants $R_0$ and $R_1$ coming from the scalar field via
\begin{align}
\label{eq:relation_BCR0R1_3d}
    BR_0 = 2CR_1\,.
\end{align}
For this particular choice, the potential also preserves the conformal invariance,
\begin{align}
    U(\phi) = \dfrac{1}{8}\lambda \phi^6\,,\quad\text{with}\quad \lambda= AR_0^2 - CR_1^2\,.
\end{align}

\paragraph{Spatial-coordinates dependent scalar field:} Now, we deal with the case $\phi = \phi(r,\theta)$. Here, for $\xi \neq 1/8$ we find
\begin{align}
    \label{eq:phi-angular-3d}
    \phi(r,\theta) = \left\{R_0 + r\left[\chi_0 \cos(\sqrt{C}\theta)  + \frac{\chi_1}{\sqrt{C}}\sin(\sqrt{C}\theta)\right]\right\}^{\frac{2\xi}{4\xi-1}}\,,
\end{align}
while $f(r)$ reads
\begin{align}
    \label{eq:f-angular-3d}
    f(r)= Ar^2 + B r^{-\frac{4\xi}{4\xi-1}} + C\,.
\end{align}
Similarly to the $D\geq 4$ case, the functional form of the metric function is preserved if $R_0=0$. Otherwise, we are restricted to demand $B=0$. In any case, the potential reads
\begin{align}
    U(\phi)&=\frac{A\xi\left(8\xi-1\right)\left(6\xi-1\right)}{(4\xi-1)^2}\phi^2+\frac{2\xi^2\left[AR_0^2+C\chi_0^2 +\chi_1^2\right]}{\left(4\xi-1\right)^2}\phi^{-\frac{2\xi-1}{\xi}}-\frac{4A\xi^2R_0\left(8\xi-1\right)}{\left(4\xi-1\right)^2}\phi^\frac{1}{2\xi}\,.
\end{align}

On the other hand, the conformal case with $\xi=1/8$ restores the quadratic polynomial \eqref{eq:f-purely-radial-xi-conf} and the relation \eqref{eq:relation_BCR0R1_3d} between the constants $B,C,R_1$ and $R_0$, so that the scalar field reads

\begin{align}
    \label{eq:phi-angular-conformal_3d}
    \phi(r,\theta) = \left\{R_0 + r\left[R_1 + \chi_0 \cos(\sqrt{C}\theta)  + \frac{\chi_1}{\sqrt{C}}\sin(\sqrt{C}\theta)\right] \right\}^{-\frac{1}{2}}\,,
\end{align}
and
\begin{align}
    \label{eq:U-conformal-3d}
    U(\phi) = \dfrac{1}{8} \left[AR_0^2 - CR_1^2 + C\chi_0^2 + \chi_1^2\right] \phi^6 \,.
\end{align}

\paragraph{Time-dependent configuration:} Finally, some comments regarding the time-dependent branch are in order. Similarly to the $D\geq 4$ case, one easily checks that a time-dependent scalar field is compatible with
\begin{equation}
f(r)=Ar^2 + Br + C\ ,
\end{equation}
with a scalar field given by,
\begin{align}\label{scalar3d}
    \phi(t,r,\theta) = \left(T(t)\sqrt{f(r)} + R_1r + R_0 + rX_1(\theta)\right)^{\frac{2\xi}{4\xi-1}}\,,
\end{align}
where,
\begin{align}
T(t)&=T_1 \cos\left(\frac{\sqrt{-\Delta}}{2}t\right) +\dfrac{T_0}{\sqrt{-\Delta}}\sin\left(\frac{\sqrt{-\Delta}}{2}t\right)\,,\quad \Delta=B^2-4AC\,.
\end{align}
The function $X_1(\theta)$ fulfills the following differential equation
\begin{align}
\dv[2]{X_1}{\theta} + C X_1 + C R_1 - \frac{1}{2}BR_0=0\,.
\end{align}
Interestingly, in contrast with the $D\geq 4$ case, here the equation $T_t^t-T_{\theta}^{\theta}=0$ forces us to choose $B=0$ unless $\xi=1/8$, without any restriction on $R_0$. For later purposes, we focus on the $C\neq 0$ case, where the scalar field reads
\begin{align}
\label{eq:scalar_field_time_dep_3d}
\begin{split}
\phi(t,r,\theta) =& \left\{R_0 + \left[ T_1 \cos\left(\frac{\sqrt{-\Delta}}{2}t\right) +\dfrac{T_0}{\sqrt{-\Delta}}\sin\left(\frac{\sqrt{-\Delta}}{2}t\right)\right] \sqrt{Ar^2+Br+C}\right. \\
&\hspace{3cm}\left. + r\left[\dfrac{BR_0}{2C} + \chi_1 \cos(\sqrt{C}\theta) + \frac{\chi_0}{\sqrt{C}}\sin(\sqrt{C}\theta)\right]\right\}^{\frac{2\xi}{4\xi-1}}\,.
\end{split}
\end{align}
Note that a real scalar field requires $A>0$ as it is the coefficient of the leading term in the asymptotic region. The self-interacting potential $U(\phi)$ is given by
\begin{align}
\label{eq:potential_3d_time_dep}
U(\phi)= \lambda_{1}\phi^2 + \lambda_2 \phi^{-\frac{2\xi-1}{\xi}} + \lambda_3 \phi^{\frac{1}{2\xi}}\,,
\end{align}
where
\begin{subequations}
\begin{align}
\label{eq:potential_3d_time_dep_lambdas}
\lambda_1 &= \frac{A\xi\left(8\xi-1\right)\left(6\xi-1\right)}{(4\xi-1)^2}\,, \\
\lambda_2 &= \dfrac{2\xi^2}{(4\xi - 1)^2}\left[ -\dfrac{1}{4}T_0^2 + \dfrac{\Delta}{4}T_1^2 + \chi_0^2 + C\chi_1^2 - \dfrac{\Delta R_0^2}{4C} \right] \,,\\
\lambda_3 &= -\frac{4A\xi^2R_0\left(8\xi-1\right)}{\left(4\xi-1\right)^2}\,.
\end{align}
\end{subequations}

Summarizing the previous sections, using this reverse approach we have solved the stealth condition $T_{\mu\nu}=0$ for $D\geq 3$, without any reference to the gravitational sector. In all the cases presented above, the stealth condition fixes the scalar field and the geometry, so the matter content codified in the stress-energy tensor dictates the form of the admissible geometries. In this sense, the scalar field does have an impact on the geometry, even when $T_{\mu\nu}=0$ is commonly interpreted as ``no backreaction''. Indeed, recognizing that the scalar field does not need to share the symmetries of the geometry significantly broadens the spectrum of non-trivial scalar profiles. In this respect, three different geometries are determined by a radial-dependent, spatial-coordinates-dependent, and time-dependent scalar field. In the latter case, a quadratic polynomial in the metric function rules the admissible geometries. Taking pertinent limits, this family includes familiar solutions such as the BTZ black hole or the (A)dS$_D$ spacetime, as we will explain in detail in the next section. On the other hand, a time-independent, axisymmetry-breaking scalar field can incorporate an extra term proportional to $r^{\frac{(D-3) - 4(D-2)\xi}{4\xi-1}}$, which is different from the usual $1/r^{(D-3)}$ decay in the metric function from GR. In this regard, it is worth emphasizing that within the non-minimal coupling $R\phi^2$, the Schwarzschild and Schwarzschild-AdS black holes are only recovered with a constant scalar field, precluding the existence of a stealth scalar field for such backgrounds.

We stress that, as mentioned in the introduction, we only have addressed point \textit{b)}. In the next section, we move to point \textit{a)} and complete the construction of a stealth solution, revisiting known results and presenting novel solutions.
\section{Applying the reverse procedure: Concrete examples}\label{Section: Applying the reverse stealth construction: Concrete examples}

In this section, we show concrete examples where the reverse approach is used. Following the path from \textit{b)} to \textit{a)}, we now delve into the search for purely gravitational theories that are compatible with the nongravitating scalar field. This problem is simplified to find gravity theories whose vacuum solutions fit in any of the cases presented in the previous sections, matching the constants from the gravity part with the ones obtained in the scalar-tensor part through this novel procedure. Through this study, we recover and generalize known cases, as well as finding novel stealth configurations. Even though we apply the reverse procedure to specific metric theories, we emphasize that the geometries determined by the stealth conditions presented in the previous section are, in principle, applicable to any other purely metric theory whose vacuum solution matches such geometries. A concrete example of these ideas is the stealth solution in AdS background (c.f. \eqref{eq:ads_metric}). Any theory that contains AdS as a vacuum solution is susceptible to being non-minimally coupled to a stealth scalar field.

We begin by conducting a literature review of all cases where the scalar-tensor sector $S_{st}[\phi,g]$ is given by \eqref{eq: Nonminimal coupling}, in the background \eqref{eq:metric_ansatz}. These works are presented in Table \ref{tab: Applications} for $S_g[g]=\int dx^D\sqrt{-g}\mathcal{L}_g$, given by the Lagrangian density $\mathcal{L}_g$, in General Relativity with and without cosmological constant~\cite{Ayon-Beato:2004nzi,Ayon-Beato:2005yoq}, New Massive Gravity (with graviton mass $m$)~\cite{Hassaine:2013cma}, Einstein--Gauss--Bonnet~\cite{BravoGaete:2013acu}, Lovelock~\cite{BravoGaete:2013djh} and Conformal Gravity~\cite{Brihaye2009,Anastasiou:2022wjq}.

\begin{center}
    \begin{tabular}{c|c|c|c}
    \hline
    $\mathcal{L}_g$ & Set-up & \shortstack{Coupling value of $\xi$} & Reference \\
    \hline\hline
    $R-2\Lambda$ & $D=3$ & Arbitrary & \cite{Ayon-Beato:2004nzi} \\ 
    \hline
    $R$ & $D\geq 3$, and $\gamma=1$ & Arbitrary & \cite{Ayon-Beato:2005yoq} \\
    \hline\rule{0pt}{4ex}    
    $\displaystyle R-2\Lambda-\frac{1}{m^2}\left(R_{\mu\nu}R^{\mu\nu}-\frac{3}{8}R^2\right)$ & $D=3$ & $1/8$ & \cite{Hassaine:2013cma} \\
    \hline\rule{0pt}{2.5ex}    
    $\displaystyle\frac{1}{2}\left(R-2\Lambda+\alpha \mathcal{G}\right)$ & $D=5$, and $\gamma=0$ & $1/6$ & \cite{BravoGaete:2013acu} \\
    \hline\rule{0pt}{4ex}
    $\displaystyle\sum_{p=0}^{n}\alpha_p\epsilon_{a_1\cdots a_D}R^{a_1 a_2}\cdots R^{a_{2p-1}a_{2p}}e^{a_{2p+1}}\cdots e^{a_D}$ & $D\geq 5$, and $\gamma=0$ & $\dfrac{(n-1)(D-1)}{4(nD-D+1)}$ & \cite{BravoGaete:2013djh} \\ \hline\rule{0pt}{2.5ex}
    $W^{\alpha\beta}_{\mu\nu}W^{\mu\nu}_{\alpha\beta}$ & $D=4$, for any $\gamma$ & $1/6$ & \cite{Brihaye2009,Anastasiou:2022wjq} \\
    \hline
    \end{tabular}
    \captionof{table}{List of works recovered by the reverse approach. Here, $m$ is the graviton mass, $n=\left[D/2\right]$, and $\alpha$ and $\alpha_p$ are the coupling constant of the Gauss-Bonnet and Lovelock densities, respectively}
    \label{tab: Applications}
\end{center}

\subsection{Stealth solutions in General Relativity}

\paragraph{Stealth over BTZ:} In three dimensions, General Relativity with negative cosmological constant admits the BTZ black hole \cite{Banados:1992wn}. The stealth overflying this black hole was first introduced in \cite{Ayon-Beato:2004nzi}, and here we revisit the solution from our approach, as a particular case of the time-dependent branch analyzed in Section \ref{Section: Three-dimensional case}. Concretely, the authors considered time and radial scalar configurations over the following metric
\begin{align}
\dd s^2 = -F(r)~\dd t^2 + \dfrac{\dd r^2}{F(r)} + r^2\left(\dd \phi - \dfrac{J}{2r^2}~\dd t \right)^2\,,
\end{align}
with $F(r)=\dfrac{r^2}{\ell^2} - M + \dfrac{J^2}{4r^2}$. In their analysis, which corresponds to the standard $\textit{a)}\rightarrow \textit{b)}$ approach under our notation, the equations to solve the stealth condition $T_{\mu\nu}=0$ force $J=0$, rendering the metric to be static, with a scalar field obeying the following profile
\begin{align}
\label{eq:btz_eloy_scalar_tr}
\phi(t,r) = \left[T(t)\sqrt{F(r)} + h(t)\right]^{\frac{2\xi}{4\xi-1}}\,. 
\end{align}
In the static case, the BTZ metric simply reads $F(r)=\dfrac{r^2}{\ell^2}-M$. From our procedure, including time-dependent scalar fields restricts the metric function to be a quadratic polynomial, namely,
\begin{align}
\phi=\phi(t,r)\,,\quad f(r)=Ar^2+Br+C\,.
\end{align}
Hence, we can link the standard and reverse procedures just by fixing,
\begin{align}\label{gluing_btz}
A=\frac{1}{\ell^2}\,,\quad  B=0\,,\quad C=-M\,.
\end{align}
At this point, we explicitly see that our constructive approach is not completed until the gravity theory is included, and gluing both parts is substantial to realize the physical relevance of the parameters or integration constants we have found. In this case, the outcome was that $A$ is a parameter related to the cosmological constant, while $-C$ represents the mass of the black hole. We stress that the scalar field itself cannot provide interpretations regarding the parameters, and there is no guarantee that the ones obtained hold arbitrarily.

We can recover the stealth scalar field from \cite{Ayon-Beato:2004nzi} as a particular case from the most general structure of the scalar fields presented in Eq.~\eqref{eq:scalar_field_time_dep_3d}, as well as the self-interacting potential~\eqref{eq:potential_3d_time_dep}-\eqref{eq:potential_3d_time_dep_lambdas}, just by applying the gluing conditions \eqref{gluing_btz} and additionally $T_0 = \chi_0 = \chi_1 =0$.

\paragraph{Stealth overflying Anti-de Sitter:} In GR, maximally symmetric spacetimes are uniquely characterized by the cosmological constant $\Lambda$. For negative values of $\Lambda$, the solution of the Einstein equations with the maximal number of symmetries is Anti-de Sitter (AdS$_{D}$), and it is described by
\begin{align}
\label{eq:ads_metric}
\dd s^2 = -\left(\frac{r^2}{\ell^2} + 1 \right)\dd t^2+ \left(\frac{r^2}{\ell^2}+1\right)^{-1}\dd r^2 + r^2\dd \Sigma_{D-2,1}^2\,,
\end{align}
with $\Lambda = -\dfrac{(D-1)(D-2)}{2\ell^2}$.
The reverse approach reduces the difficulty of finding stealth solutions by linking the geometries. This is why it is easy to obtain time-dependent stealth scalar fields over AdS$_D$ for any dimension $D\geq 3$. 
As remarked in the previous section, time-dependent scalar fields constrain the metric function to be a quadratic polynomial. For $D=3$, the stealth solution is of the form,
\begin{align}
\phi=\phi(t,r,\theta)\,,\quad f(r)=Ar^2+Br+C\,.
\end{align}
Clearly, we can link our procedure to the AdS$_3$ spacetime by fixing
\begin{align}\label{gluing_ads}
A=\frac{1}{\ell^2}\,,\quad B=0\,,\quad  C=1\,.
\end{align}
Applying these conditions to \eqref{eq:scalar_field_time_dep_3d}, a novel solution is obtained. This is,
\begin{align}
\phi(t,r,\theta) = \left[R_0 + T_1\cos\left(\frac{t}{\ell}\right)\sqrt{\frac{r^2}{\ell^2}+1} + \chi_1\cos(\theta) \right]^{\frac{2\xi}{4\xi-1}}\,,
\end{align}
after absorbing integration constants $T_0$ and $\chi_0$ due to the isometries of the metric, while the potential is obtained from ~\eqref{eq:potential_3d_time_dep}-\eqref{eq:potential_3d_time_dep_lambdas}. The stealth solution over (A)dS$_{D}$ can be obtained in $D\geq4$ just by considering \eqref{gluing_ads} in the solutions provided in Section \ref{Section: The reverse stealth construction}. Finally, our procedure also recovers the Minkowski stealth from Ref.~\cite{Ayon-Beato:2005yoq}, by considering $A=B=0$ and $C=1$ in the time-dependent branch \eqref{eq: Branches}.

It is noteworthy that the reverse approach allows coupling a stealth scalar field in a non-minimal manner to any metric theory that admits a maximally symmetric spacetime as a solution.

\subsection{Stealth solutions beyond General Relativity}\label{Subsection: Revisiting and finding novel stealths from the literature}

\paragraph{New Massive Gravity:} A stealth solution from Ref.~\cite{Hassaine:2013cma} was found in New Massive Gravity, which is also a sensitive metric theory that can be coupled to a stealth scalar field. This means that we consider the following action principle,
\begin{align}
S[g,\phi]=\int d^3x\sqrt{-g}\left[R-2\Lambda-\frac{1}{m^2}\left(R_{\mu\nu}R^{\mu\nu}-\frac{3}{8}R^2\right)-\left(\frac{1}{2}\nabla_\mu\phi\nabla^\mu\phi+\frac{1}{2}\xi R\phi^2+U\left(\phi\right)\right)\right]\ .
\end{align}
Then, the corresponding stealth solution belongs to the radial dependent configuration in $D=3$ presented in Section \ref{Section: Three-dimensional case}. In the conformal case, $\xi = 1/8$, we have 
\begin{align}
\phi=\phi(r)\,,\quad f(r)=Ar^2+Br+C\,.
\end{align}
Hence, the stealth solution from the above reference can be successfully retrieved by fixing
\begin{align}\label{gluing_nmg}
A=\frac{1}{\ell^2}\,,\quad  B=b\,,\quad C=-4GM\,,
\end{align}
and the conformal potential
\begin{equation}\label{pot_nmg}
U(\phi)=\lambda\phi^6\,.
\end{equation}
From the reverse approach, when the scalar field depends on time, the metric function also behaves as a quadratic polynomial. This simple observation and the linear decay in the vacuum solution of New Massive Gravity make it possible to present an extension of the stealth from Ref.~\cite{Hassaine:2013cma}. Concretely, using
\begin{align}
\phi=\phi(t,r)\,,\quad f(r)=Ar^2+Br+C\,,
\end{align}
and the same conditions \eqref{gluing_nmg}, the novel time-dependent scalar field reads
\begin{align}
\phi(t,r) = \left[ T_1\cosh\left(\dfrac{\Delta}{2}t\right)\sqrt{\frac{r^2}{\ell^2} + br - 4GM} + R_0 - \frac{bR_0}{8GM}r \right]^{-\frac{1}{2}}\,,
\end{align}
where $\Delta = b^2 +\frac{16GM}{\ell^2}$, and we have already absorbed $T_0$ by time translations invariance. This time-dependent configuration demands the relation for the parameter of the potential
\begin{align}
\lambda = \dfrac{\Delta}{4}T_1^2 - \dfrac{\Delta R_0^2}{4C}\,,
\end{align}
which is fixed without variation. When $T_1\to 0$, we recover the purely-radial stealth from Ref.~\cite{Hassaine:2013cma} previously mentioned.

\paragraph{Gauss-Bonnet and Lovelock:} Lovelock gravity \cite{Lovelock:1971yv} is the most general $D-$dimensional gravity model with second order equations. It coincides with GR in three and four dimensions. The $n-$th member of the Lovelock family is a topological invariant in $D=2n$, so they modify GR in $D\geq 2n+1$ dimensions. To apply the reverse procedure to this theory, we coupled the scalar field in the following form,
\begin{equation}
\begin{aligned}
  S[\phi,g] &= \int d^Dx\sqrt{-g}\sum_{p=0}^{n} \alpha_p\delta^{\beta_1 \cdots \beta_{2 p}}{ }_{\gamma_1 \cdots \gamma_{2 p}} R^{\gamma_1 \gamma_2}{ }_{\beta_1 \beta_2} \cdots R^{\gamma_{2 p-1} \gamma_{2 p}}{ }_{\beta_{2 p}-\beta_{2 p}}\\
  &\quad -\int d^Dx\sqrt{-g}\left(\frac{1}{2}\nabla_\mu\phi\nabla^\mu\phi+\frac{1}{2}\xi R\phi^2+U\left(\phi\right)\right)\ , \quad 1 \leq n \leq\left[\frac{D-1}{2}\right]\ .
  \end{aligned}
\end{equation}

In Lovelock gravities, maximally symmetric solutions demand some relations between the coupling constants $\alpha_p$ of the curvature terms. Under the hypothesis that the theory admits a unique AdS-vacua, black hole solutions have been found in Ref. \cite{Crisostomo:2000bb}, and the metric function is given by
\begin{align}
	\label{eq:f_Black_hole_scan}
		f(r) = \dfrac{r^2}{\ell^2} + Br^{-\frac{D-2n-1}{n}} + \gamma\,.
\end{align} 
The shape of the metric function allows us to enhance these black hole solutions with a scalar field depending on the transverse coordinates, as shown in Section \ref{Subsubsection: Transverse sector}. This is,
\begin{align}
\phi=\phi(r,\vec{x})\,,\quad f(r)=Ar^2+Br^{\frac{(D-3)-4(D-2)\xi}{4\xi-1}}+\gamma\,,
\end{align}
recovers the metric function \eqref{eq:f_Black_hole_scan} after the following conditions are applied
\begin{align}\label{gluing_lovelock}
	A=\dfrac{1}{\ell^2}\,,\quad \xi= \dfrac{(n-1)(D-1)}{4(nD-D+1)}\,.
\end{align}
The integration constant $B$ is related to the mass of the black hole. Stealths in Gauss-Bonnet and Lovelock gravities were found a few years ago~\cite{BravoGaete:2013acu, BravoGaete:2013djh}, where the authors restrict the analysis to flat manifolds ($\gamma=0$) only. Our approach not only confirms their result but also generalizes it to base manifolds with non-vanishing curvature and to spatial-coordinates dependent scalar fields.

\paragraph{Conformal gravity:} Like the Schwarzschild black hole in General Relativity, Birkhoff's theorem yields the Riegert black hole in conformal gravity, whose stealth was originally found in \cite{Brihaye2009}, and reconsidered recently as part of a scalar-tensor renormalization program in \cite{Anastasiou:2022wjq}. Following the notation of the latter, for $\xi=1/6$ the solution is given by,
\begin{align}
\phi=\phi(r)\,,\quad f(r)= k+ \dfrac{6mG}{r_0} - \dfrac{2}{r_0}\left(k+\dfrac{3mG}{r_0}\right)r - \dfrac{2mG}{r} - \dfrac{\lambda r^2}{3}\,.
\end{align}
It is clear that from our reverse procedure, this solution is recovered by the four-dimensional time-independent branch with a metric function \eqref{eq: f conformal}, a potential with the form presented in \eqref{eq: U conformal}, and a purely radial scalar field given in \eqref{radial_scalar}. This is achieved by imposing the conditions,
\begin{align}
A&= -\dfrac{\lambda}{3} - \dfrac{2mG}{r_0^3}\,,&B&= \dfrac{2mG \sigma^3}{r_0^3}\,, &R_1&= \dfrac{1}{\sigma}\,, &R_0&= -\dfrac{r_0}{\sigma}\,, &\sigma&=\sqrt{-\dfrac{k+\frac{2mG}{r_0}+\frac{\lambda r_0^2}{3}}{2\nu}}\,,
\end{align}
where $m$, $\lambda$ and $r_0$ are integration constants, whereas $\nu$ is a parameter that modules the behavior of the associated potential.

\paragraph{Quasitopological gravities:} Although they do not possess a topological origin as Lovelock's model, Quasitopological Gravities~\cite{Oliva:2010eb,Myers:2010ru,Dehghani:2011vu,Cisterna:2017umf} are constructed in $D\geq 5$ upon $n-$th order curvature interactions. Unlike Lovelock, the value of $n$ in Quasitopological Gravities is not restricted by the spacetime dimension and can grow arbitrarily\footnote{Recently, it was shown that in $D \geq 5$, the infinite tower of Quasitopological Gravities regularizes the curvature singularity of the Schwarzschild black hole~\cite{Bueno:2024dgm}.}. These theories emerge as gravitational models with second-order Euler-Lagrange equations when evaluated precisely in constant curvature internal spaces, as in our ansatz \eqref{eq:metric_ansatz}, and they satisfy a $n-$th order algebraic equation for the metric function $f(r)$~\cite{Bueno:2019ycr,Bueno:2022res}. It was shown that the $n-$th Quasitopological gravity fulfills 
\begin{align*}
\label{eq: metric function QTG}
    \dfrac{D-2}{2}r^{D-2n-1}\left[\gamma-f(r)\right]^n = M\,,
\end{align*}
where $M$ is an integration constant associated with the ADM mass of the solution. For that reason, the reverse procedure allows straightforwardly accommodating a novel stealth solution considering the spatial-coordinates dependent scalar field \eqref{scalar_rx},
\begin{align}
\phi=\phi(r,\vec{x})\,,\quad f(r)=Ar^2+Br^{\frac{(D-3)-4(D-2)\xi}{4\xi-1}}+\gamma\,,
\end{align}
by fixing $\xi$ as
\begin{align}
	\xi = \dfrac{(n-1)(D-1)}{4(nD-D+1)}\,,
\end{align}
and the self-interacting potential as shown in \eqref{eq: potencial_general_structure}.

To end this section, we remark on some general features. As shown through the previous examples and as emphasized in the introduction, the approach \textit{a)} to \textit{b)} is not bijective. Furthermore, the reverse construction does not impose restrictions on the purely gravitational sector (i.e., the left-hand side of \eqref{eq: Stealth condition}). For example, if $\xi =\xi_{D-1}=\dfrac{D-3}{4(D-2)}$, the metric function acquires the form
\begin{align}
    f(r)=Ar^2-M\,,
\end{align}
which we recognize as a non-rotating BTZ-like black hole~\cite{Banados:1992wn}. Thus, any gravitational theory whose vacuum solution admits such a profile is compatible with stealth, with the aforementioned coupling constant.

In sum, we retrieved and generalized known stealth solutions from the literature. Judging by the form of the metric function \eqref{eq: Metric function angular} in the time-independent branch, the values of $\xi$ and $B$ become significant, with the former controlling the decay of the metric function and the latter being an integration constant that cannot be interpreted solely by solving $T_{\mu\nu}=0$. Therefore, the physical interpretation of such constants is understood once the gravitational theory is specified.

\section{Thermodynamic imprints of stealth black holes}\label{Section: Thermodynamic imprints of stealth black holes}
It is well known that gravitational perturbations on the stealth background can backreact, and in consequence, at the perturbative level, the presence of this ``gravitationally undetectable'' object is exposed. Consequently, perturbations departing from the stealth configuration manifest clear differences with those departing from its trivial counterpart~\cite{Faraoni:2010mj}. However, what is less known and has not yet been explored is whether these differences persist at the thermodynamic level, allowing one to distinguish the stealth black hole from the trivial one. The spectrum of stealth solutions obtained by the reverse procedure leaves us in a good position for these purposes. This section aims to show that, indeed, there are differences induced by the presence of a stealth scalar field exposed in the thermodynamic quantities.

Let us choose, as an exemplary case, the five-dimensional Einstein-Gauss-Bonnet action as the purely metric theory $S_g[g]$, endowed with a non-minimally coupled scalar field in $S_{st}[\phi,g]$. This is,
\begin{equation}\label{EsGB}
S[\phi,g]=\int d^5 x \sqrt{-g}\left[\frac{1}{2\kappa}\left(R-2 \Lambda+\alpha\mathcal{G}\right)-\frac{1}{2} \nabla_\mu \phi \nabla^\mu \phi-\frac{\xi}{2} R \phi^2-U(\phi)\right]\,,
\end{equation}
where the term $\mathcal{G}$ stand for the Gauss-Bonnet invariant given by $\mathcal{G}=R^2-4 R_{\alpha \beta} R^{\alpha \beta}+R_{\alpha \beta \mu \nu} R^{\alpha \beta \mu \nu}$, the self-interacting potential is given by,
\begin{equation}
U(\phi)=\frac{1}{6\ell^2}\phi^2\ ,
\end{equation}
whose stealth solution was found in \cite{BravoGaete:2013acu}, and it reads,
\begin{align}\label{solthermo}
\phi(r)=\frac{\phi_0}{r}\,,\quad f(r)=\frac{r^2}{\ell^2}-M\,.
\end{align}
The reverse procedure recovers this solution by setting $D=5$ and $n=2$ in \eqref{gluing_lovelock}.

The Euclidean approach in the calculation of the thermodynamical quantities establishes that the partition function for a thermodynamical ensemble is identified with the Euclidean path integral in the saddlepoint approximation around the classical Euclidean solution~\cite{PhysRevD.15.2752}. Once the symmetries of the spacetimes are known, Palais’ “principle of symmetric criticality” states that the extremum of the action can be found imposing such symmetries on a minisuperspace metric and then varying the action. These symmetries extremely simplify the Euclidean procedure, as the general Hamiltonian formalism is not always available. However, this principle is not universal, as the symmetries imposed on the minisuperpsace must not over-reduce the degrees of freedom in the action, otherwise the set of field equations emerging from the reduced action will not correspond to the covariant field equations. Accordingly, the Euclidean continuation of this solution reads,
\begin{equation}
\dd s_E^2=N(r)f(r)\dd\tau^2+\frac{dr^2}{f(r)}+r^2(\dd x_1^2+\dd x_2^2+\dd x_3^2)\ ,
\end{equation}
with $\phi=\phi(r)$. The flat coordinates span $x_i\in(0,\beta_i)$ with $i=\{1,2,3\}$.
The period $\beta$ of the Euclidean time $\tau$ is related to the temperature $T$ of the black hole by $\beta=1/T$. In consequence, the reduced Euclidean action $S_E$ can be written in Hamiltonian form as follows,
\begin{equation}
S_E=\beta \sigma \int_{r_{+}}^{\infty}d r N\mathcal{H}+B_E\ ,
\end{equation}
where $B_E$ is a boundary term introduced in action \eqref{EsGB} and $\sigma=\beta_1\beta_2\beta_3$ is the volume of the base manifold. The variation of $B_E$ cancels out all the contributions coming from variations of the bulk action, defining a well-posed variational principle to have an extremum on the classical solution. The reduced constraint is given by,
\begin{align}
\begin{split}
\mathcal{H}=&-\frac{3r}{2\kappa}(2 f+r f')-\frac{1}{2}r \delta^{ij}\partial_i \phi \partial_j \phi-r^3 \left[\frac{1}{2}f(\partial_r\phi)^2+U(\phi)+\frac{\Lambda}{\kappa}\right]+6\alpha f f'\\
&\hspace{2cm}+\xi \left\{\frac{3}{2}r\phi^2(rf'+2 f)+2r\delta^{ij}(\partial_i\phi\partial_j\phi+\phi\partial_i\partial_j\phi)+\sqrt{f}\partial_r\left[\sqrt{f}r^3\left(\phi^2\right)'\right]\right\}\, .
\end{split}
\end{align}
It is reassuring to check that, despite $\mathcal{H}$ being an involved expression, the following off-shell relations are satisfied,
\begin{equation}\label{rels}
\begin{aligned}
\frac{\delta S_E}{\delta N}&=-2\sqrt{g}\frac{\mathcal{E}_{tt}}{{N^3f}}\,,\\
\frac{\delta S_E}{\delta f}&=\sqrt{g}\left(\mathcal{E}_{rr}-\frac{\mathcal{E}_{tt}}{N^2f^2}\right)\,,\\
\frac{\delta S_E}{\delta\phi }&=\sqrt{g}\mathcal{E}_{\phi}\,,
\end{aligned}
\end{equation}
confirming the validity of the boundary term. Its variation is given by,
\begin{align}
\begin{split}
    \delta B_E&=\beta\sigma\bigg\{-\frac{r^2}{2 \kappa}\left(3 \delta f+2\kappa r\phi' f\delta \phi \right) + \frac{1}{2} r^2 \xi\left[2r\phi'(2 f \delta \phi+\phi\delta f) + \phi\left(4 r f \delta \phi' + 3 \phi \delta f \right) \right.\\
    &\hspace{6cm}\left.-2 r \phi f'\delta \phi + 6 \alpha f\delta f \right]\bigg\}\bigg\rvert_{r_+}^{\infty}
\end{split}
\end{align}
The variation of the fields at the horizon $r_+$ acquires the following form,
\begin{align}
\left.\delta f\right|_{r_{+}} =-\left.f^{\prime}\right|_{r_{+}} \delta r_{+}\,,\quad \left.\delta \phi\right|_{r_{+}}=\delta \phi\left(r_{+}\right)-\left.\phi^{\prime}\right|_{r_{+}} \delta r_{+}\,.
\end{align}
Then, the variation of the boundary term at the horizon is
\begin{equation}
\delta B_E(r_+)=\delta\left\{\frac{2\pi}{\kappa}\sigma r_+^3\left[1-\kappa\xi\phi(r_+)^2\right]\right\}\,,
\end{equation}
which can be readily integrated as
\begin{equation}\label{bhor}
B_E(r_+)=\frac{A_+}{4 \tilde{G}_{+}}\quad\text{with}\,,\quad \tilde{G}_{+}=\frac{G}{\left(1-\kappa\xi\phi\left(r_{+}\right)^2\right)}\,,
\end{equation}
where $A_+=\sigma r_+^3$. On the other hand, the variations of the fields at infinity are,
\begin{align}
\left.\delta f\right|_{\infty}=-\delta M\,,\quad \left.\delta\phi\right|_{\infty}=\frac{\delta \phi_0}{r}\,.
\end{align}
With these quantities, the variation of the boundary term at infinity is,
\begin{equation}\label{binf}
\delta B_E(\infty)=\frac{3 r^2}{2\ell^2\kappa}\beta\sigma(\ell^2-4\alpha\kappa)\delta M+\beta\sigma\left(6\alpha M-\frac{1}{12}\phi_0^2\right) \delta M-\frac{1}{3} \beta \sigma M \phi_0 \delta\phi_0\ .
\end{equation}
There is a potential quadratic divergence in the first term of this expression. However, solution \eqref{solthermo} requires that $\ell^2=4\kappa\alpha$ (the Chern--Simons point), which ensures the finiteness of the variation of the boundary term \eqref{binf}. This is,
\begin{equation}\label{binf_reg}
\delta B_E(\infty)=\beta\sigma\left(\frac{3\ell^2}{2\kappa} M-\frac{1}{12}\phi_0^2\right) \delta M-\frac{1}{3} \beta \sigma M \phi_0 \delta\phi_0\ .
\end{equation}

Observe that $\delta B_E(\infty)$ is not necessarily a closed 1-form. Condition $\delta^2B_E\left(\infty\right)=0$ leads,
\begin{align}
    \phi_0 \delta\phi_0\wedge \delta M =0\,.
\end{align}
This equation raises two possibilities: $\phi_0=0$ or $\delta\phi_0\wedge \delta M=0$. The former switches off the scalar field, while the latter allows the possibility of computing the mass for the stealth solution. This implies to consider $\phi_0$ as a function of $M$, and the variation $\delta B_E(\infty)$ is now exact by Poincar\'e's lemma. Since we are working in the grand canonical ensemble, the temperature is fixed, and the boundary term at infinity can be integrated as,
\begin{align}
\label{eq: Boundary term}
 B_E(\infty) = \beta \sigma\int \dd M \left[\frac{3\ell^2}{2\kappa} M-\frac{1}{12}\phi_0^2(M)-\frac{1}{3} M \phi_0(M) \pdv{\phi_0(M)}{M}\right]\,.
\end{align}
It is remarkable to see that the value of the Euclidean action is finite for any differentiable function $\phi_0(M)$ (see below for concrete examples), in which case,
\begin{equation}
S_E=B(\infty)-B(r_+)=\beta \sigma\int \dd M \left[\frac{3\ell^2}{2\kappa} M-\frac{1}{12}\phi_0^2(M)-\frac{1}{3} M \phi_0(M) \pdv{\phi_0(M)}{M}\right]-\frac{A_+}{4 \tilde{G}_{+}}\ ,
\end{equation}
up to an arbitrary constant without variation. Since the Gibbs free energy $F$ is related to the Euclidean action by $S_E=\beta F=\beta \mathcal{M}-\mathcal{S}$, the mass $\mathcal{M}$ and the entropy $\mathcal{S}$ are given by the
standard thermodynamical relations
\begin{align}
\label{thermo}
\mathcal{M}&=\frac{\partial S_E}{\partial\beta}=\sigma\frac{3\ell^2}{4\kappa} M^2+\Delta \mathcal{M}\,,\\
\mathcal{S}&=\left(\beta\frac{\partial}{\partial\beta}-1\right)S_E=\frac{A_+}{4 \tilde{G}_{+}}\,.
\end{align}
It is clear from \eqref{thermo} that the shift in the mass with respect to the trivial configuration is given by,
\begin{equation}\label{DeltaM}
\Delta \mathcal{M}=\sigma\int \dd M \left[-\frac{1}{12}\phi_0^2(M)-\frac{1}{3} M \phi_0(M) \pdv{\phi_0(M)}{M}\right]\ ,
\end{equation}
and the value of the mass is altered by different choices of $\phi_0$ as a function of $M$. This is precisely the case for a hairy black hole with secondary hair. Stealth solutions with primary hair are excluded from a thermodynamic perspective as their mass is not integrable.

In general, the shift in the mass generated by the stealth field must fulfill two conditions: It has to be non-trivial, meaning that $\delta\left(\Delta \mathcal{M}\right) \neq 0$; and non-divergent, which in this case is already guaranteed by construction for any differential function $\phi_0(M)$, according to formula \eqref{eq: Boundary term}. In this regard, consider the following examples.

In the case that the scalar field is rigid, $\partial\phi_0\left(M\right)/\partial M = 0$, the mass reads
\begin{align}
    \mathcal{M} = \sigma \left(\dfrac{3\ell^2}{4\kappa}M^2 - \dfrac{1}{12}\phi_0^2 M\right)\,,
\end{align}
fulfilling the above criteria. A healthy and non-trivial example can be achieved by $\phi_0(M)=M\ell$, obtaining
\begin{align}
    \mathcal{M} = \sigma\left(\dfrac{3\ell^2}{4\kappa} M^2 - \dfrac{5\ell^2}{36}M^3\right)\,. 
\end{align}
It is interesting to analyze some extreme cases at this point. It could be the case that the shift in the mass is tuned in such a way that it cancels out the black hole's mass. To achieve this, and according to \eqref{DeltaM}, it is straightforward to see that
\begin{equation}
\mathcal{M}=0\ ,
\end{equation}
is achieved by choosing $\phi_0(M) = \sqrt{\dfrac{6M \ell^2}{\kappa}}$. This kind of black hole with vanishing mass is not new. Black hole configurations with zero mass have already been reported in \cite{Cai:2009ac,Anabalon:2011bw,Correa:2013bza}. However, the novelty of our result relies on the fact that the black hole mass is screened by a non-gravitating source.


\section{Summary and outlook}\label{Section: Summary and outlook}

In this article, we presented a novel method for constructing stealth solutions using a reverse approach. The latter proposes a generic procedure to construct stealth solutions by restricting the spacetime geometry only by imposing the stealth conditions $T_{\mu\nu}=0$. As opposed to the standard procedure of obtaining stealth solutions — from step \textit{a)} to step \textit{b)} — starting with step \textit{b)} by solving $T_{\mu\nu}=0$ leads to a wide range of geometries compatible with stealth configurations. Then, if such geometries match the vacuum solutions of some gravity theory, such theory can be coupled to a non-gravitating scalar field, completing step \textit{a)}. Another consequence of this approach is that the role of the integration constants in the conserved charges may not be immediately clear. This is expected, as their nature relies on the specific theory to which the stealth solution is coupled. To ground this approach in a specific scalar-tensor theory, we focused on the non-minimally coupled scalar field in $D$ dimensions.

As we show in Section \ref{Section: The reverse stealth construction} and \ref{Section: Three-dimensional case}, the reverse approach makes it evident that the coordinate dependence of the scalar profile controls the form of the metric function in the homogeneous ansatz. For a time-dependent scalar field, the metric function is determined by $f(r)=Ar^2+Br+C$. In contrast, in the time-independent case, there are two possible scenarios for the metric function: A spatial-coordinates dependent $\phi(r,\vec{x})$ and a purely radial $\phi(r)$ scalar field. The spatial-coordinates dependent scalar field is compatible with a metric function $f(r)=Ar^2+Br^{\frac{(D-3) - 4(D-2)\xi}{4\xi-1}}+C$ exhibiting a term whose decay depends on the coupling parameter $\xi$. This term is inherently different from the term found in pure GR, which is proportional to $1/r^{D-3}$ and incompatible with a stealth solution as it is obtained only when $\xi=0$, yielding a constant scalar field. For the purely radial stealth scalar field, the metric function admits an involved expression of Gauss’ hypergeometric and Gamma functions, which, in the conformal case $\xi=\xi_D$, reduces to $f(r)=Ar^2+2\gamma R_1 r/R_0+\gamma+B\left(R_1r+R_0\right)^{D-1}/r^{D-3}$.

In this work, the inclusion of a self-interacting potential in the scalar-tensor sector $S_{st}[\phi,g]$ of the action that supports stealth configurations takes the form
\begin{align}
    U(\phi)= \lambda_1\phi^2+\lambda_2\phi^{-\frac{2\xi-1}{\xi}}+\lambda_3\phi^{\frac{1}{2\xi}}\,.
\end{align}
This potential consists of three terms: one massive $\propto \phi^2$ and two terms involving powers of $\xi$. The massive term of the potential provides the vacuum of the theory, given by (A)dS or Minkowski spacetimes. The second term $\propto\phi^{-\frac{2\xi-1}{\xi}}$ emerges as an extension of the conformal sector, in the sense that if $\xi=\xi_D$, the associated power reduces to the conformal case. Lastly, the term $\propto \phi^{\frac{1}{2\xi}}$ induces pure deviations from the conformally coupled scalar field. Its effect is controlled by a parameter that modulates the affine profile of the scalar field.

The reverse approach was applied to some concrete examples analyzed in Section \ref{Section: Applying the reverse stealth construction: Concrete examples}, along with its thermodynamic implications as seen in Section \ref{Section: Thermodynamic imprints of stealth black holes}. Naturally, all the known stealth solutions with non-minimally coupled scalar fields were recovered, and additionally, in some cases, they were generalized. This is because, in this reverse procedure, the problem finally reduces to match the parameters and integration constants in the stealth solution with those generated from the purely metric side. For instance, if one fixes the coupling parameter to its conformal value in one dimension lower, $\xi=\xi_{D-1} = \dfrac{D-3}{4(D-2)}$, the metric function \eqref{eq: Metric function angular} acquires a non-rotating BTZ-like form
\begin{align}\label{btz-like}
	f(r)= Ar^2-M\,,
\end{align}
irrespective of the gravity model under consideration. This means that not only
\begin{align}
    S[g,\phi] = S_g[g]-\int d^Dx\sqrt{-g}\left[\frac{1}{2}\nabla_\mu\phi\nabla^\mu\phi+\frac{1}{2}\xi_{D-1} R\phi^2+U\left(\phi\right) \right]\,,
\end{align}
with $S_g[g]=\dfrac{R-2\Lambda}{2\kappa}$ admits a stealth solution as shown in \cite{Ayon-Beato:2004nzi}, but also any $D-$dimensional gravity theory $S_g[g]$ whose family of solutions contains a BTZ-like metric function \eqref{btz-like} and in particular AdS$_D$ spacetimes. Furthermore, we have shown that the reverse approach paves a direct path to finding new non-gravitating configurations. Concretely, novel stealth solutions have been found in New Massive Gravity, Lovelock, and Quasitopological Gravities (see Section \ref{Subsection: Revisiting and finding novel stealths from the literature}).

It is important to note that even when the stress-energy tensor vanishes, the stealth solution can still exhibit tangible physical effects. With the solutions provided by the reverse procedure, we focused on studying the thermodynamic imprints left by the non-gravitating configurations. To develop this point, the five-dimensional Einstein-Gauss-Bonnet action with a non-minimally coupled scalar field is considered in Section \ref{Section: Thermodynamic imprints of stealth black holes}. Surprisingly, the black hole mass is screened by the non-gravitating source. Consequently, a stealth scalar field is not only detectable at the perturbative level but also at the thermodynamic level. This result is physically relevant as it may have phenomenological implications: If we measure the mass of a celestial body with our current tools, we would not be able to distinguish whether the measurable mass includes a contribution from a stealth scalar field.

As shown in this article, the reverse approach only requires a $\st$ action and a particular ansatz for the geometry. Here, we explored the non-minimal coupling $R\phi^2$ over a static and spherically symmetric ansatz. In this regard, it is possible to go beyond the aforementioned non-minimal coupling towards, for example, $G^{\mu\nu}\nabla_\mu\phi\nabla_\nu\phi$, where it is known that a stealth configuration $\phi=\phi(t,r)$ is compatible with the Schwarzschild solution~\cite{Babichev:2013cya}. Motivated by the existence of this solution, it is feasible to explore the reverse approach within this framework. On the other hand, it is natural to consider an anisotropic ansatz, where stealth solutions are already known on a Lifshitz-type background~\cite{Ayon-Beato:2015qfa}. For such an ansatz, the reverse approach could unveil black hole solutions with (bulk or asymptotic) anisotropic scaling. 

In this work, we have seen that the stealth scalar field has a physical effect on the mass. Another avenue to pursue is to consider a rotating ansatz and explore the consequences of a stealth configuration on the angular momentum. More generally, one can naturally ask if some electric or color charge is changed under the presence of a stealth configuration. The latter may be achieved through a charged non-minimal coupling.



\section*{Acknowledgments}
The authors thank E. Ayon-Beato, F. Cánfora, C. Corral, D. Flores-Alfonso, O. Fuentealba, M. Hassaïne, R. Troncoso, and J. Zanelli for useful discussions. L.G. and K.L. especially thank C. Henriquez-Baez for her contribution during the first stage of this project. C.E. is funded by Agencia Nacional de Investigacion y Desarrollo (ANID) through Proyecto FONDECYT Iniciación folio 11221063, Etapa 2024. The work of K.L. is supported by FONDECYT grant 3230618. 

\appendix 

\section{Case $\xi=1/4$}\label{Appendix: Section: Case xi=1/4}

This appendix deals with the case $\xi=1/4$ that we omitted in the main article. Starting from the metric ansatz \eqref{eq:metric_ansatz}, the off-diagonal components of the $T_{\mu}^{\nu}$ tensor are simultaneously solved only in the case when the scalar field is separable, and it does not depend on the transverse sector, namely
\begin{align}
\phi(t,r,x_i) = T(t)R(r)\,.
\end{align}
Furthermore, the $T_{t}^r$ equation imposes an additional constraint, proportional to
\begin{align}
    \dot{T}f' = 0\,,
\end{align}
which forces purely-radial scalar fields or metrics with $f(r)=cst$. In any case, we automatically have $T^{1}_{1}=T^{2}_{2}=\dots=T^{{D-2}}_{{D-2}}$ for the spatial sector, meaning that we have three effective equations to solve: $T_t^t=0$, $T_r^r=0$, and $T_i^{i}=0$.

In the first case, when $\phi=\phi(r)$ is a purely-radial scalar field, the combination $T_t^t - T_r^r=0$ turns out to be a differential equation for $\phi(r)$, which is solved by
\begin{align}
    \phi(r)=R_0e^{R_1r}\,,
\end{align}
but in this case, $f(r)$ is generically given by an elliptic integral. 

Regarding the second case, when $\phi = \phi(t,r)$ forces the metric function to be a constant. In this case, we can only dress the Minkowski spacetime, obtained when $f(r)=\gamma=1$. The scalar field reads
\begin{align}
\label{ap: scalar t-r 1/4}
    \phi(t,r)=R_0 e^{R_1(r^2-t^2)}\,,
\end{align}
and the potential $U(\phi)$ yields
\begin{align}
    \label{ap:potencial 1/4}
    U(\phi) = \left[\dfrac{(D-1)R_1}{R_0} + 2\ln\left(\dfrac{\phi}{R_0}\right)\right]\phi^2\,,
\end{align}
in agreement with the results of Ref.~\cite{Ayon-Beato:2005yoq}.

\section{Explicit form of $T^t_t-T^i_i$}\label{Appendix: Explicit form}

In this appendix, we show the subtraction $T^t_t-T^i_i$ for a general ansatz of the form
\begin{align}
	\label{ap: eq: metric ansatz general}
	ds^2=-f(r)dt^2+\frac{dr^2}{f(r)}+h^2(r)d\Sigma^2_{D-2,\gamma}\,,
\end{align}
with $h(r)$ an arbitrary function, where
\begin{align}
	d\Sigma_{D-2,\gamma}^2 =\Omega\left(\vec{x}\right)^2 dx_i dx^i\,,\quad \Omega\left(\vec{x}\right)=\frac{1}{1+\frac{\gamma}{4}x_i x^i}\,.
\end{align}
The construction of Section \ref{Section: The reverse stealth construction} is recovered for the particular case $h(r)=r$.

The difference $T^t_t-T^i_i$ is
\begin{align}
	T^t_t-T^{i}_{i}&=\dfrac{ \psi^{\frac{1}{4\xi-1}}\Omega\left(\vec{x}\right)}{8\sqrt{f}h^2(4\xi-1)}\left( \Xi_2 \sum_{i=1}^{D-2} x_i^2 + \Xi_1 \sum_{i=1}^{D-2}b_ix_i + \Xi_0 \right) = 0\,,
\end{align}
where $\Xi_{2,1,0}$ are, in general, differential equations for $T$, $h$ and $f$. Concretely, $\Xi_2$ is
\begin{subequations}
    \label{ap: eq: Xis}
    \begin{align}
        \begin{split}
            \Xi_2\equiv & 8k\xi^2 h^2 \ddot{T} + \Big\{\gamma\xi (8\xi - 2) f^2hh'' + 2(D-3)\gamma\xi(4\xi - 1)f^2{h'}^2 \\ 
		&\hspace{3cm}+ \gamma\xi \left[4(5-D)\xi + D - 4\right]ff'hh' + \left.  \left[\gamma\xi (1-4\xi)ff'' - 2\gamma \xi^2 {f'}^2 \right] h^2 \right. \\
		&\hspace{3cm}+ 2(D-3)\gamma^2\xi(1-4\xi)f\Big\} T+ 2\xi(4\xi-1)(4ah + \gamma R)f^{3/2} hh''  \\ 
		&\hspace{3cm}+ 8(D-3)\xi  \left[\frac{4(D-2)\xi - D +3}{D-3}h + \dfrac{\gamma(4\xi-1)}{4}R \right]f^{3/2}{h'}^{2}\\ 
		&\hspace{0.5cm}+ \left\{ 4a\xi\left[D-4 - 4(D-3)\xi\right]\sqrt{f}f'h^2 + \left[\gamma(D-4)(1-4\xi)\sqrt{f}f'R + 8k\xi^2f^{3/2}R'\right]h \right\}h'\\
		&\hspace{1cm}+ 4a\xi(1-4\xi)\sqrt{f}f''h^3+ (\gamma\xi(1-4\xi)\sqrt{f}f''R- 4k\xi^2\sqrt{f}f'R')h^2 + 4k\xi\bigg\{2(D-3)a\\
		&\hspace{3cm}+\xi\left[\gamma \sum_{i=1}^{D-2} c_i -  (8D-20)a\right]\bigg\}\sqrt{f}h + 2(D-3)\gamma^2\xi(1-4\xi)\sqrt{f}R\Biggr\}\,,
        \end{split}
    \end{align}
whereas $\Xi_1$,
\begin{align}
    \begin{split}
        \Xi_1 \equiv & 8\xi(4\xi-1)f^{3/2}h^2h'' + 8\xi\left[ 4(D-2)\xi - D+3 \right]f^{3/2}h{h'}^{2} -  4\xi\left[4(D-3)\xi - D+4\right]\sqrt{f}f'h^2h' \\ 
	&\hspace{3cm} + 4\xi(1-4\xi)\sqrt{f}f''h^3 - 8\gamma \xi(4(D-2)\xi - D+3)\sqrt{f}h \,,
    \end{split}
\end{align}
and finally, 
\begin{align}
	\begin{split}
		\Xi_0\equiv & ~32\xi^2h^2\ddot{T} + \Big\{8\xi(4\xi-1)f^2hh'' + 8(D-3)\xi(4\xi-1)f^2{h'}^2 + 4\xi\left[4(5-D)\xi + D - 4\right]ff'hh'  \\
		& + \left[ -4\xi(4\xi-1)ff'' - 8\xi^2{f'}^2 \right]h^2 - 8(D-3)\gamma\xi(4\xi-1)f \Big\}T + 8\xi(4\xi-1)f^{3/2}hh''\times \\
		&\nspace\times \left(R + h\sum_{i=1}^{D-2}c_i\right) + 32(D-2)\xi \left[ \left(\xi - \dfrac{D-3}{4(D-2)}\right)h\sum_{i=1}^{D-2}c_i + \dfrac{D-3}{4(D-2)}(4\xi-1)R\right]f^{3/2}{h'}^2 \\ 
		& + \Bigg\{ -16\xi(D-3)\sum_{i=1}^{D-2}c_i \left[ \xi - \dfrac{D-4}{4(D-3)} \right]\sqrt{f}f'h^2 + \left[ -4\xi(D-4)(4\xi-1)\sqrt{f}f'R \right. \\
		& \left.+ 32\xi^2 f^{3/2}R' \right]h  \Bigg\} h'- 4\xi(4\xi-1)\sum_{i=1}^{D-2}c_i\sqrt{f}f''h^3 + \left(-4\xi(4\xi-1)\sqrt{f}f''R \right. \\ 
		&\left.-16\xi^2\sqrt{f}f'R'\right)h^2 + \xi\left[-16(2D-5)\left(\xi - \dfrac{2(D-3)}{8D-20}\right) \sum_{i=2}^{D-2}c_i\gamma + 64\xi a\right]\sqrt{f}h \\
		&\hspace{6cm}- 8(D-3)\gamma\xi (4\xi-1)\sqrt{f}R\,.
	\end{split}
\end{align}
\end{subequations}

Particularly, for the time-independent scalar fields of Section \ref{Subsection: Time-independent configuration Dgeq4}, the discussion of Eq.~\eqref{eq: Ttt-Tii} follows by imposing $T(t)=0$, $R(r)=R_1r+R_0$, and $g(r)=r$.

\clearpage

\printbibliography

@article{Langlois:2015cwa,
	Archiveprefix = {arXiv},
	Author = {Langlois, David and Noui, Karim},
	Doi = {10.1088/1475-7516/2016/02/034},
	Eprint = {1510.06930},
	Journal = {JCAP},
	Pages = {034},
	Primaryclass = {gr-qc},
	Title = {{Degenerate higher derivative theories beyond Horndeski: evading the Ostrogradski instability}},
	Volume = {02},
	Year = {2016}}

@inproceedings{Langlois:2017mdk,
	Archiveprefix = {arXiv},
	Author = {Langlois, David},
	Booktitle = {{52nd Rencontres de Moriond on Gravitation}},
	Eprint = {1707.03625},
	Pages = {221--228},
	Primaryclass = {gr-qc},
	Title = {{Degenerate Higher-Order Scalar-Tensor (DHOST) theories}},
	Year = {2017}}

@article{Langlois:2018dxi,
	Archiveprefix = {arXiv},
	Author = {Langlois, David},
	Doi = {10.1142/S0218271819420069},
	Eprint = {1811.06271},
	Journal = {Int. J. Mod. Phys. D},
	Number = {05},
	Pages = {1942006},
	Primaryclass = {gr-qc},
	Title = {{Dark energy and modified gravity in degenerate higher-order scalar\textendash{}tensor (DHOST) theories: A review}},
	Volume = {28},
	Year = {2019}}

@article{Kobayashi:2019hrl,
	Archiveprefix = {arXiv},
	Author = {Kobayashi, Tsutomu},
	Doi = {10.1088/1361-6633/ab2429},
	Eprint = {1901.07183},
	Journal = {Rept. Prog. Phys.},
	Number = {8},
	Pages = {086901},
	Primaryclass = {gr-qc},
	Reportnumber = {RUP-19-3},
	Title = {{Horndeski theory and beyond: a review}},
	Volume = {82},
	Year = {2019}}

@article{Gleyzes:2014dya,
	Archiveprefix = {arXiv},
	Author = {Gleyzes, J\'er\^ome and Langlois, David and Piazza, Federico and Vernizzi, Filippo},
	Doi = {10.1103/PhysRevLett.114.211101},
	Eprint = {1404.6495},
	Journal = {Phys. Rev. Lett.},
	Number = {21},
	Pages = {211101},
	Primaryclass = {hep-th},
	Title = {{Healthy theories beyond Horndeski}},
	Volume = {114},
	Year = {2015}}

@article{Crisostomi:2016tcp,
	Archiveprefix = {arXiv},
	Author = {Crisostomi, Marco and Hull, Matthew and Koyama, Kazuya and Tasinato, Gianmassimo},
	Doi = {10.1088/1475-7516/2016/03/038},
	Eprint = {1601.04658},
	Journal = {JCAP},
	Pages = {038},
	Primaryclass = {hep-th},
	Title = {{Horndeski: beyond, or not beyond?}},
	Volume = {03},
	Year = {2016}}

@article{Horndeski:1974wa,
	Author = {Horndeski, Gregory Walter},
	Doi = {10.1007/BF01807638},
	Journal = {Int. J. Theor. Phys.},
	Pages = {363--384},
	Title = {{Second-order scalar-tensor field equations in a four-dimensional space}},
	Volume = {10},
	Year = {1974}}

@article{kerr,
	Author = {Kerr, Roy P.},
	Doi = {10.1103/PhysRevLett.11.237},
	Issue = {5},
	Journal = {Phys. Rev. Lett.},
	Month = {Sep},
	Numpages = {0},
	Pages = {237--238},
	Publisher = {American Physical Society},
	Title = {Gravitational Field of a Spinning Mass as an Example of Algebraically Special Metrics},
	Url = {https://link.aps.org/doi/10.1103/PhysRevLett.11.237},
	Volume = {11},
	Year = {1963}}

@article{BenAchour:2018dap,
	Archiveprefix = {arXiv},
	Author = {Ben Achour, Jibril and Liu, Hongguang},
	Doi = {10.1103/PhysRevD.99.064042},
	Eprint = {1811.05369},
	Journal = {Phys. Rev. D},
	Number = {6},
	Pages = {064042},
	Primaryclass = {gr-qc},
	Title = {{Hairy Schwarzschild-(A)dS black hole solutions in degenerate higher order scalar-tensor theories beyond shift symmetry}},
	Volume = {99},
	Year = {2019}}

@article{Motohashi:2018wdq,
	Archiveprefix = {arXiv},
	Author = {Motohashi, Hayato and Minamitsuji, Masato},
	Doi = {10.1016/j.physletb.2018.04.041},
	Eprint = {1804.01731},
	Journal = {Phys. Lett. B},
	Pages = {728--734},
	Primaryclass = {gr-qc},
	Reportnumber = {YITP-18-27},
	Title = {{General Relativity solutions in modified gravity}},
	Volume = {781},
	Year = {2018}}

@article{Babichev:2017guv,
	Archiveprefix = {arXiv},
	Author = {Babichev, Eugeny and Charmousis, Christos and Leh\'ebel, Antoine},
	Doi = {10.1088/1475-7516/2017/04/027},
	Eprint = {1702.01938},
	Journal = {JCAP},
	Pages = {027},
	Primaryclass = {gr-qc},
	Reportnumber = {LPT-ORSAY-17-03},
	Title = {{Asymptotically flat black holes in Horndeski theory and beyond}},
	Volume = {04},
	Year = {2017}}

@article{Charmousis:2014zaa,
	Archiveprefix = {arXiv},
	Author = {Charmousis, Christos and Kolyvaris, Theodoros and Papantonopoulos, Eleftherios and Tsoukalas, Minas},
	Doi = {10.1007/JHEP07(2014)085},
	Eprint = {1404.1024},
	Journal = {JHEP},
	Pages = {085},
	Primaryclass = {gr-qc},
	Title = {{Black Holes in Bi-scalar Extensions of Horndeski Theories}},
	Volume = {07},
	Year = {2014}}

@article{Bocharova,
	Author = {N. Bocharova and K. Bronnikov and V. Melnikov},
	Journal = {Vestn. Mosk. Univ. Fiz. Astron.},
	Pages = {706},
	Title = {An exact solution of the system of Einstein equations and mass-free scalar field},
	Volume = {7},
	Year = {1970}}

@article{janis,
	Author = {Janis, Allen I. and Newman, Ezra T. and Winicour, Jeffrey},
	Doi = {10.1103/PhysRevLett.20.878},
	Issue = {16},
	Journal = {Phys. Rev. Lett.},
	Month = {Apr},
	Numpages = {0},
	Pages = {878--880},
	Publisher = {American Physical Society},
	Title = {Reality of the Schwarzschild Singularity},
	Url = {https://link.aps.org/doi/10.1103/PhysRevLett.20.878},
	Volume = {20},
	Year = {1968}}

@article{MaselliBHs,
	Author = {Maselli, Andrea and Silva, Hector O. and Minamitsuji, Masato and Berti, Emanuele},
	Doi = {10.1103/PhysRevD.92.104049},
	Issue = {10},
	Journal = {Phys. Rev. D},
	Month = {Nov},
	Numpages = {12},
	Pages = {104049},
	Publisher = {American Physical Society},
	Title = {Slowly rotating black hole solutions in Horndeski gravity},
	Url = {https://link.aps.org/doi/10.1103/PhysRevD.92.104049},
	Volume = {92},
	Year = {2015}}

@article{Hui,
	Archiveprefix = {arXiv},
	Author = {Hui, Lam and Nicolis, Alberto},
	Doi = {10.1103/PhysRevLett.110.241104},
	Eprint = {1202.1296},
	Journal = {Phys. Rev. Lett.},
	Pages = {241104},
	Primaryclass = {hep-th},
	Slaccitation = {%%CITATION = ARXIV:1202.1296;%%},
	Title = {{No-Hair Theorem for the Galileon}},
	Volume = {110},
	Year = {2013}}

@article{SotiriouBHs,
	Archiveprefix = {arXiv},
	Author = {Sotiriou, Thomas P.},
	Doi = {10.1088/0264-9381/32/21/214002},
	Eprint = {1505.00248},
	Journal = {Class. Quant. Grav.},
	Number = {21},
	Pages = {214002},
	Primaryclass = {gr-qc},
	Title = {{Black Holes and Scalar Fields}},
	Volume = {32},
	Year = {2015}}

@article{heusler,
	Author = {Chru{\'s}ciel, Piotr T. and Costa, Jo{\~a}o Lopes and Heusler, Markus},
	Doi = {10.12942/lrr-2012-7},
	Issn = {1433-8351},
	Journal = {Living Reviews in Relativity},
	Month = {May},
	Number = {1},
	Publisher = {Springer Science and Business Media LLC},
	Title = {Stationary Black Holes: Uniqueness and Beyond},
	Url = {http://dx.doi.org/10.12942/lrr-2012-7},
	Volume = {15},
	Year = {2012}}

@article{israel,
	Author = {Israel, Werner},
	Doi = {10.1103/PhysRev.164.1776},
	Issue = {5},
	Journal = {Phys. Rev.},
	Month = {Dec},
	Numpages = {0},
	Pages = {1776--1779},
	Publisher = {American Physical Society},
	Title = {Event Horizons in Static Vacuum Space-Times},
	Url = {https://link.aps.org/doi/10.1103/PhysRev.164.1776},
	Volume = {164},
	Year = {1967}}

@article{wheeler,
	Author = {Ruffini, Remo and Wheeler, John A.},
	Doi = {10.1063/1.3022513},
	Journal = {Phys. Today},
	Number = {1},
	Pages = {30},
	Title = {{Introducing the black hole}},
	Volume = {24},
	Year = {1971}}

@article{Brown:1986nw,
    author = "Brown, J. David and Henneaux, M.",
    title = "{Central Charges in the Canonical Realization of Asymptotic Symmetries: An Example from Three-Dimensional Gravity}",
    doi = "10.1007/BF01211590",
    journal = "Commun. Math. Phys.",
    volume = "104",
    pages = "207--226",
    year = "1986"
}

@article{Henneaux:2002wm,
	Archiveprefix = {arXiv},
	Author = {Henneaux, Marc and Martinez, Cristian and Troncoso, Ricardo and Zanelli, Jorge},
	Doi = {10.1103/PhysRevD.65.104007},
	Eprint = {hep-th/0201170},
	Journal = {Phys. Rev. D},
	Pages = {104007},
	Reportnumber = {CECS-PHY-01-09, ULB-TH-02-07},
	Title = {{Black holes and asymptotics of 2+1 gravity coupled to a scalar field}},
	Volume = {65},
	Year = {2002}}

@article{Babichev:2023rhn,
	Archiveprefix = {arXiv},
	Author = {Babichev, Eugeny and Charmousis, Christos and Hassaine, Mokhtar and Lecoeur, Nicolas},
	Doi = {10.1103/PhysRevD.107.084050},
	Eprint = {2302.02920},
	Journal = {Phys. Rev. D},
	Number = {8},
	Pages = {084050},
	Primaryclass = {gr-qc},
	Title = {{Conformally coupled scalar in Lovelock theory}},
	Volume = {107},
	Year = {2023}}

@article{Crisostomo:2000bb,
	Archiveprefix = {arXiv},
	Author = {Crisostomo, Juan and Troncoso, Ricardo and Zanelli, Jorge},
	Doi = {10.1103/PhysRevD.62.084013},
	Eprint = {hep-th/0003271},
	Journal = {Phys. Rev. D},
	Pages = {084013},
	Reportnumber = {CECS-PHY-00-01, ULB-TH-00-01},
	Title = {{Black hole scan}},
	Volume = {62},
	Year = {2000}}

@article{BravoGaete:2013acu,
	Archiveprefix = {arXiv},
	Author = {Bravo Gaete, Moises and Hassaine, Mokhtar},
	Doi = {10.1103/PhysRevD.88.104011},
	Eprint = {1308.3076},
	Journal = {Phys. Rev. D},
	Pages = {104011},
	Primaryclass = {hep-th},
	Title = {{Topological black holes for Einstein-Gauss-Bonnet gravity with a nonminimal scalar field}},
	Volume = {88},
	Year = {2013}}

@article{BravoGaete:2013djh,
	Archiveprefix = {arXiv},
	Author = {Bravo Gaete, Mois\'es and Hassaine, Mokhtar},
	Doi = {10.1007/JHEP11(2013)177},
	Eprint = {1309.3338},
	Journal = {JHEP},
	Pages = {177},
	Primaryclass = {hep-th},
	Title = {{Planar AdS black holes in Lovelock gravity with a nonminimal scalar field}},
	Volume = {11},
	Year = {2013}}

@article{Oliva:2010eb,
	Archiveprefix = {arXiv},
	Author = {Oliva, Julio and Ray, Sourya},
	Doi = {10.1088/0264-9381/27/22/225002},
	Eprint = {1003.4773},
	Journal = {Class. Quant. Grav.},
	Pages = {225002},
	Primaryclass = {gr-qc},
	Reportnumber = {CECS-PHY-10-03},
	Title = {{A new cubic theory of gravity in five dimensions: Black hole, Birkhoff's theorem and C-function}},
	Volume = {27},
	Year = {2010}}

@article{Myers:2010ru,
	Archiveprefix = {arXiv},
	Author = {Myers, Robert C. and Robinson, Brandon},
	Doi = {10.1007/JHEP08(2010)067},
	Eprint = {1003.5357},
	Journal = {JHEP},
	Pages = {067},
	Primaryclass = {gr-qc},
	Title = {{Black Holes in Quasi-topological Gravity}},
	Volume = {08},
	Year = {2010}}

@article{Dehghani:2011vu,
	Archiveprefix = {arXiv},
	Author = {Dehghani, M. H. and Bazrafshan, A. and Mann, R. B. and Mehdizadeh, M. R. and Ghanaatian, M. and Vahidinia, M. H.},
	Doi = {10.1103/PhysRevD.85.104009},
	Eprint = {1109.4708},
	Journal = {Phys. Rev. D},
	Pages = {104009},
	Primaryclass = {hep-th},
	Title = {{Black Holes in Quartic Quasitopological Gravity}},
	Volume = {85},
	Year = {2012}}

@article{Cisterna:2017umf,
	Archiveprefix = {arXiv},
	Author = {Cisterna, Adolfo and Guajardo, Luis and Hassaine, Mokhtar and Oliva, Julio},
	Doi = {10.1007/JHEP04(2017)066},
	Eprint = {1702.04676},
	Journal = {JHEP},
	Pages = {066},
	Primaryclass = {hep-th},
	Title = {{Quintic quasi-topological gravity}},
	Volume = {04},
	Year = {2017}}

@article{Bueno:2019ycr,
	Archiveprefix = {arXiv},
	Author = {Bueno, Pablo and Cano, Pablo A. and Hennigar, Robie A.},
	Doi = {10.1088/1361-6382/ab5410},
	Eprint = {1909.07983},
	Journal = {Class. Quant. Grav.},
	Number = {1},
	Pages = {015002},
	Primaryclass = {hep-th},
	Reportnumber = {IFT-UAM/CSIC-19-124},
	Title = {{(Generalized) quasi-topological gravities at all orders}},
	Volume = {37},
	Year = {2020}}

@article{Bueno:2022res,
	Archiveprefix = {arXiv},
	Author = {Bueno, Pablo and Cano, Pablo A. and Hennigar, Robie A. and Lu, Mengqi and Moreno, Javier},
	Doi = {10.1088/1361-6382/aca236},
	Eprint = {2203.05589},
	Journal = {Class. Quant. Grav.},
	Number = {1},
	Pages = {015004},
	Primaryclass = {hep-th},
	Reportnumber = {CERN-TH-2022-038},
	Title = {{Generalized quasi-topological gravities: the whole shebang}},
	Volume = {40},
	Year = {2023}}

@article{Bueno:2024dgm,
	Archiveprefix = {arXiv},
	Author = {Bueno, Pablo and Cano, Pablo A. and Hennigar, Robie A.},
	Eprint = {2403.04827},
	Month = {3},
	Primaryclass = {gr-qc},
	Title = {{Regular Black Holes From Pure Gravity}},
	Year = {2024}}

@article{Ayon-Beato:2004nzi,
	Archiveprefix = {arXiv},
	Author = {Ayon-Beato, Eloy and Martinez, Cristian and Zanelli, Jorge},
	Doi = {10.1007/s10714-005-0213-x},
	Eprint = {hep-th/0403228},
	Journal = {Gen. Rel. Grav.},
	Pages = {145--152},
	Reportnumber = {CECS-PHY-04-04},
	Title = {{Stealth scalar field overflying a (2+1) black hole}},
	Volume = {38},
	Year = {2006}}

@article{Ayon-Beato:2005yoq,
	Archiveprefix = {arXiv},
	Author = {Ayon-Beato, Eloy and Martinez, Cristian and Troncoso, Ricardo and Zanelli, Jorge},
	Doi = {10.1103/PhysRevD.71.104037},
	Eprint = {hep-th/0505086},
	Journal = {Phys. Rev. D},
	Pages = {104037},
	Reportnumber = {CECS-PHY-05-05},
	Title = {{Gravitational Cheshire effect: Nonminimally coupled scalar fields may not curve spacetime}},
	Volume = {71},
	Year = {2005}}

@book{Wheeler:1998vs,
	Author = {Wheeler, J. A. and Ford, K.},
	Title = {{Geons, black holes, and quantum foam: A life in physics}},
	Year = {1998}}

@article{Babichev:2013cya,
	Archiveprefix = {arXiv},
	Author = {Babichev, Eugeny and Charmousis, Christos},
	Doi = {10.1007/JHEP08(2014)106},
	Eprint = {1312.3204},
	Journal = {JHEP},
	Pages = {106},
	Primaryclass = {gr-qc},
	Reportnumber = {LPT-ORSAY-13-105},
	Title = {{Dressing a black hole with a time-dependent Galileon}},
	Volume = {08},
	Year = {2014}}

@article{Minamitsuji:2018vuw,
	Archiveprefix = {arXiv},
	Author = {Minamitsuji, Masato and Motohashi, Hayato},
	Doi = {10.1103/PhysRevD.98.084027},
	Eprint = {1809.06611},
	Journal = {Phys. Rev. D},
	Number = {8},
	Pages = {084027},
	Primaryclass = {gr-qc},
	Reportnumber = {YITP-18-104},
	Title = {{Stealth Schwarzschild solution in shift symmetry breaking theories}},
	Volume = {98},
	Year = {2018}}

@article{Charmousis:2019vnf,
	Archiveprefix = {arXiv},
	Author = {Charmousis, Christos and Crisostomi, Marco and Gregory, Ruth and Stergioulas, Nikolaos},
	Doi = {10.1103/PhysRevD.100.084020},
	Eprint = {1903.05519},
	Journal = {Phys. Rev. D},
	Number = {8},
	Pages = {084020},
	Primaryclass = {hep-th},
	Title = {{Rotating Black Holes in Higher Order Gravity}},
	Volume = {100},
	Year = {2019}}

@article{Hassaine:2006gz,
	Archiveprefix = {arXiv},
	Author = {Hassaine, Mokhtar},
	Doi = {10.1088/0305-4470/39/27/008},
	Eprint = {hep-th/0606159},
	Journal = {J. Phys. A},
	Pages = {8675--8680},
	Reportnumber = {CECS-PHY-06-15},
	Title = {{Analogies between self-duality and stealth matter source}},
	Volume = {39},
	Year = {2006}}

@article{Anabalon:2009qt,
	Archiveprefix = {arXiv},
	Author = {Anabalon, Andres and Maeda, Hideki},
	Doi = {10.1103/PhysRevD.81.041501},
	Eprint = {0907.0219},
	Journal = {Phys. Rev. D},
	Pages = {041501},
	Primaryclass = {hep-th},
	Reportnumber = {AEI-2009-059, CECS-PHY-09-05},
	Title = {{New Charged Black Holes with Conformal Scalar Hair}},
	Volume = {81},
	Year = {2010}}

@article{Faraoni:2010mj,
	Archiveprefix = {arXiv},
	Author = {Faraoni, Valerio and Moreno, Andres F. Zambrano},
	Doi = {10.1103/PhysRevD.81.124050},
	Eprint = {1006.1936},
	Journal = {Phys. Rev. D},
	Pages = {124050},
	Primaryclass = {gr-qc},
	Title = {{Are stealth scalar fields stable?}},
	Volume = {81},
	Year = {2010}}

@article{Maeda:2012tu,
	Archiveprefix = {arXiv},
	Author = {Maeda, Hideki and Maeda, Kei-ichi},
	Doi = {10.1103/PhysRevD.86.124045},
	Eprint = {1208.5777},
	Journal = {Phys. Rev. D},
	Pages = {124045},
	Primaryclass = {gr-qc},
	Reportnumber = {CECS-PHY-12-06},
	Title = {{Creation of the universe with a stealth scalar field}},
	Volume = {86},
	Year = {2012}}

@article{Caldarelli:2013gqa,
	Archiveprefix = {arXiv},
	Author = {Caldarelli, Marco M. and Charmousis, Christos and Hassaine, Mokhtar},
	Doi = {10.1007/JHEP10(2013)015},
	Eprint = {1307.5063},
	Journal = {JHEP},
	Pages = {015},
	Primaryclass = {hep-th},
	Title = {{AdS black holes with arbitrary scalar coupling}},
	Volume = {10},
	Year = {2013}}

@article{Ayon-Beato:2013bsa,
	Archiveprefix = {arXiv},
	Author = {Ayon-Beato, Eloy and Garc\'ia, Alberto A. and Ram\'irez-Baca, P. Isaac and Terrero-Escalante, C\'esar A.},
	Doi = {10.1103/PhysRevD.88.063523},
	Eprint = {1307.6534},
	Journal = {Phys. Rev. D},
	Number = {6},
	Pages = {063523},
	Primaryclass = {gr-qc},
	Title = {{Conformal stealth for any standard cosmology}},
	Volume = {88},
	Year = {2013}}

@article{Hassaine:2013cma,
	Archiveprefix = {arXiv},
	Author = {Hassaine, Mokhtar},
	Doi = {10.1103/PhysRevD.89.044009},
	Eprint = {1311.4623},
	Journal = {Phys. Rev. D},
	Number = {4},
	Pages = {044009},
	Primaryclass = {hep-th},
	Title = {{Rotating AdS black hole stealth solution in D=3 dimensions}},
	Volume = {89},
	Year = {2014}}

@article{Cardenas:2014kaa,
	Archiveprefix = {arXiv},
	Author = {Cardenas, Marcela and Fuentealba, Oscar and Martinez, Cristi\'an},
	Doi = {10.1103/PhysRevD.90.124072},
	Eprint = {1408.1401},
	Journal = {Phys. Rev. D},
	Number = {12},
	Pages = {124072},
	Primaryclass = {hep-th},
	Title = {{Three-dimensional black holes with conformally coupled scalar and gauge fields}},
	Volume = {90},
	Year = {2014}}

@article{Ayon-Beato:2015qfa,
	Archiveprefix = {arXiv},
	Author = {Ay\'on-Beato, Eloy and Hassaine, Mokhtar and Ju\'arez-Aubry, Mar\'ia Montserrat},
	Eprint = {1506.03545},
	Month = {6},
	Primaryclass = {gr-qc},
	Title = {{Stealths on Anisotropic Holographic Backgrounds}},
	Year = {2015}}

@article{Ayon-Beato:2015mxf,
	Archiveprefix = {arXiv},
	Author = {Ayon-Beato, Eloy and Ram\'irez-Baca, P. Isaac and Terrero-Escalante, C\'esar A.},
	Doi = {10.1103/PhysRevD.97.043505},
	Eprint = {1512.09375},
	Journal = {Phys. Rev. D},
	Number = {4},
	Pages = {043505},
	Primaryclass = {gr-qc},
	Title = {{Cosmological stealths with nonconformal couplings}},
	Volume = {97},
	Year = {2018}}

@inproceedings{Henneaux:2015tar,
	Archiveprefix = {arXiv},
	Author = {Henneaux, Marc and P\'erez, Alfredo and Tempo, David and Troncoso, Ricardo},
	Booktitle = {{International Workshop on Higher Spin Gauge Theories}},
	Doi = {10.1142/9789813144101-0009},
	Eprint = {1512.08603},
	Pages = {139--157},
	Primaryclass = {hep-th},
	Reportnumber = {CECS-PHY-15-12},
	Title = {{Extended anti-de Sitter Hypergravity in $2+1$ Dimensions and Hypersymmetry Bounds}},
	Year = {2017}}

@article{Cisterna:2016nwq,
	Archiveprefix = {arXiv},
	Author = {Cisterna, Adolfo and Hassaine, Mokhtar and Oliva, Julio and Rinaldi, Massimiliano},
	Doi = {10.1103/PhysRevD.94.104039},
	Eprint = {1609.03430},
	Journal = {Phys. Rev. D},
	Number = {10},
	Pages = {104039},
	Primaryclass = {gr-qc},
	Title = {{Static and rotating solutions for Vector-Galileon theories}},
	Volume = {94},
	Year = {2016}}

@article{Smolic:2017bic,
	Archiveprefix = {arXiv},
	Author = {Smoli\'c, Ivica},
	Doi = {10.1103/PhysRevD.97.084041},
	Eprint = {1711.07490},
	Journal = {Phys. Rev. D},
	Number = {8},
	Pages = {084041},
	Primaryclass = {gr-qc},
	Reportnumber = {ZTF-EP-17-11},
	Title = {{Spacetimes dressed with stealth electromagnetic fields}},
	Volume = {97},
	Year = {2018}}

@article{Aviles:2018vnf,
	Archiveprefix = {arXiv},
	Author = {Avil\'es, Luis and Maeda, Hideki and Martinez, Cristian},
	Doi = {10.1088/1361-6382/aaea9f},
	Eprint = {1808.10040},
	Journal = {Class. Quant. Grav.},
	Number = {24},
	Pages = {245001},
	Primaryclass = {gr-qc},
	Title = {{Exact black-hole formation with a conformally coupled scalar field in three dimensions}},
	Volume = {35},
	Year = {2018}}

@article{Quinzacara:2019pes,
	Archiveprefix = {arXiv},
	Author = {Quinzacara, Cristian and Meza, Paola and Sampson, Almeira and Valenzuela, Mauricio},
	Doi = {10.1088/1475-7516/2023/03/032},
	Eprint = {1904.00494},
	Journal = {JCAP},
	Pages = {032},
	Primaryclass = {hep-th},
	Title = {{Electromagnetically and gravitationally stealth fields}},
	Volume = {03},
	Year = {2023}}

@article{Bravo-Gaete:2021kgt,
	Archiveprefix = {arXiv},
	Author = {Bravo-Gaete, Moises and Juarez-Aubry, Maria Montserrat and Velazquez-Rodriguez, Gerardo},
	Doi = {10.1103/PhysRevD.105.084009},
	Eprint = {2112.01483},
	Journal = {Phys. Rev. D},
	Number = {8},
	Pages = {084009},
	Primaryclass = {hep-th},
	Title = {{Lifshitz black holes in four-dimensional critical gravity}},
	Volume = {105},
	Year = {2022}}

@article{Anastasiou:2022wjq,
	Archiveprefix = {arXiv},
	Author = {Anastasiou, Giorgos and Araya, Ignacio J. and Busnego-Barrientos, Mairym and Corral, Cristobal and Merino, Nelson},
	Doi = {10.1103/PhysRevD.107.104049},
	Eprint = {2212.04364},
	Journal = {Phys. Rev. D},
	Number = {10},
	Pages = {104049},
	Primaryclass = {hep-th},
	Title = {{Conformal renormalization of scalar-tensor theories}},
	Volume = {107},
	Year = {2023}}

@article{Flores-Alfonso:2023fkd,
	Archiveprefix = {arXiv},
	Author = {Flores-Alfonso, Daniel and Lopez-Monsalvo, Cesar S. and Maceda, Marco},
	Eprint = {2311.12985},
	Month = {11},
	Primaryclass = {gr-qc},
	Title = {{Gravitational Waves from Thurston Geometries}},
	Year = {2023}}

@article{Banados:1992wn,
    author = "Banados, Maximo and Teitelboim, Claudio and Zanelli, Jorge",
    title = "{The Black hole in three-dimensional space-time}",
    eprint = "hep-th/9204099",
    archivePrefix = "arXiv",
    reportNumber = "PRINT-92-0151 (CHILE), IASSNS-HEP-92-29",
    doi = "10.1103/PhysRevLett.69.1849",
    journal = "Phys. Rev. Lett.",
    volume = "69",
    pages = "1849--1851",
    year = "1992"
}

@article{Lovelock:1971yv,
    author = "Lovelock, D.",
    title = "{The Einstein tensor and its generalizations}",
    doi = "10.1063/1.1665613",
    journal = "J. Math. Phys.",
    volume = "12",
    pages = "498--501",
    year = "1971"
}

@article{Deffayet1,
	Author = {C. Deffayet and G. Esposito-Farese and A. Vikman},
	Doi = {10.1103/PhysRevD.79.084003},
	Issue = {8},
	Journal = {Phys. Rev. D},
	Month = {Apr},
	Numpages = {6},
	Pages = {084003},
	Publisher = {American Physical Society},
	Title = {Covariant Galileon},
	Url = {http://link.aps.org/doi/10.1103/PhysRevD.79.084003},
	Volume = {79},
	Year = {2009}}

@article{Deffayet2,
	Author = {C. Deffayet and G. Esposito-Farese and A. Vikman},
	Doi = {10.1103/PhysRevD.80.064015},
	Issue = {6},
	Journal = {Phys. Rev. D},
	Month = {Sep},
	Numpages = {5},
	Pages = {064015},
	Publisher = {American Physical Society},
	Title = {Generalized Galileons: All scalar models whose curved background extensions maintain second-order field equations and stress tensors},
	Url = {http://link.aps.org/doi/10.1103/PhysRevD.80.064015},
	Volume = {80},
	Year = {2009}}

@article{Kobayashi,
	Archiveprefix = {arXiv},
	Author = {Kobayashi, Tsutomu and Yamaguchi, Masahide and Yokoyama, Jun'ichi},
	Doi = {10.1143/PTP.126.511},
	Eprint = {1105.5723},
	Journal = {Prog. Theor. Phys.},
	Pages = {511-529},
	Primaryclass = {hep-th},
	Reportnumber = {KUNS-2339, RESCEU-9-11},
	Slaccitation = {%%CITATION = ARXIV:1105.5723;%%},
	Title = {{Generalized G-inflation: Inflation with the most general second-order field equations}},
	Volume = {126},
	Year = {2011}}

@article{Rinaldi,
	Archiveprefix = {arXiv},
	Author = {Rinaldi, Massimiliano},
	Doi = {10.1103/PhysRevD.86.084048},
	Eprint = {1208.0103},
	Journal = {Phys. Rev. D},
	Pages = {084048},
	Primaryclass = {gr-qc},
	Slaccitation = {%%CITATION = ARXIV:1208.0103;%%},
	Title = {{Black holes with non-minimal derivative coupling}},
	Volume = {86},
	Year = {2012}}

@article{Ayon-Beato:2024xgp,
    author = "Ay\'on-Beato, Eloy and Hassaine, Mokhtar and S\'anchez, Pedro A.",
    title = "{Non-Noetherian Conformal Cheshire Effect}",
    eprint = "2408.00086",
    archivePrefix = "arXiv",
    primaryClass = "hep-th",
    month = "7",
    year = "2024"
}

@article{Brihaye2009,
  title = {Spherical structures in conformal gravity and its scalar-tensor extension},
  author = {Brihaye, Y. and Verbin, Y.},
  journal = {Phys. Rev. D},
  volume = {80},
  issue = {12},
  pages = {124048},
  numpages = {17},
  year = {2009},
  month = {Dec},
  publisher = {American Physical Society},
  doi = {10.1103/PhysRevD.80.124048},
  url = {https://link.aps.org/doi/10.1103/PhysRevD.80.124048}
}

@article{PhysRevD.15.2752,
  title = {Action integrals and partition functions in quantum gravity},
  author = {Gibbons, G. W. and Hawking, S. W.},
  journal = {Phys. Rev. D},
  volume = {15},
  issue = {10},
  pages = {2752--2756},
  numpages = {0},
  year = {1977},
  month = {May},
  publisher = {American Physical Society},
  doi = {10.1103/PhysRevD.15.2752},
  url = {https://link.aps.org/doi/10.1103/PhysRevD.15.2752}
}

@article{Correa:2013bza,
    author = "Correa, Francisco and Hassaine, Mokhtar",
    title = "{Thermodynamics of Lovelock black holes with a nonminimal scalar field}",
    eprint = "1312.4516",
    archivePrefix = "arXiv",
    primaryClass = "hep-th",
    doi = "10.1007/JHEP02(2014)014",
    journal = "JHEP",
    volume = "02",
    pages = "014",
    year = "2014"
}

@article{Anabalon:2011bw,
    author = "Anabalon, Andres and Canfora, Fabrizio and Giacomini, Alex and Oliva, Julio",
    title = "{Black holes with gravitational hair in higher dimensions}",
    eprint = "1108.1476",
    archivePrefix = "arXiv",
    primaryClass = "hep-th",
    reportNumber = "AEI-2011-053",
    doi = "10.1103/PhysRevD.84.084015",
    journal = "Phys. Rev. D",
    volume = "84",
    pages = "084015",
    year = "2011"
}

@article{Cai:2009ac,
    author = "Cai, Rong-Gen and Liu, Yan and Sun, Ya-Wen",
    title = "{A Lifshitz Black Hole in Four Dimensional R**2 Gravity}",
    eprint = "0909.2807",
    archivePrefix = "arXiv",
    primaryClass = "hep-th",
    doi = "10.1088/1126-6708/2009/10/080",
    journal = "JHEP",
    volume = "10",
    pages = "080",
    year = "2009"
}

\end{document}